\documentclass[aps,pre,reprint,superscriptaddress]{revtex4-1}

\usepackage{graphicx}
\usepackage[english]{babel} 
\usepackage{amsfonts}

\begin{document}

\title{Cause and Cure of Sloppiness in Ordinary Differential Equation Models}

\author{Christian T\"{o}nsing}
\email[christian.toensing@fdm.uni-freiburg.de]{}
\affiliation{Institute of Physics, University of Freiburg, 79104 Freiburg, Germany}
\author{Jens Timmer}
\affiliation{Institute of Physics, University of Freiburg, 79104 Freiburg, Germany}
\affiliation{BIOSS Centre for Biological Signalling Studies, University of Freiburg, 79104 Freiburg, Germany}
\author{Clemens Kreutz}
\affiliation{Institute of Physics, University of Freiburg, 79104 Freiburg, Germany}

\begin{abstract}
Data-based mathematical modeling of biochemical reaction networks, e.g. by nonlinear ordinary differential equation (ODE) models, has been successfully applied. In this context, parameter estimation and uncertainty analysis is a major task in order to assess the quality of the description of the system by the model. Recently, a broadened eigenvalue spectrum of the Hessian matrix of the objective function covering orders of magnitudes was observed and has been termed as sloppiness. In this work, we investigate the origin of sloppiness from structures in the sensitivity matrix arising from the properties of the model topology and the experimental design. Furthermore, we present strategies using optimal experimental design methods in order to circumvent the sloppiness issue and present non-sloppy designs for a benchmark model.\end{abstract}

\maketitle

\section{Introduction}

Mathematical modeling of biochemical processes became increasingly popular during the last decade and resulted in new interdisciplinary scientific disciplines like Systems Biology. A major goal is the establishment of mathematical models describing cellular processes at a molecular level \cite{kitano_systems_2002}. Frequently, ordinary differential equations (ODE) are used to mathematically represent the dynamic behavior of cellular components \cite{wolkenhauer_systems_2008}. To be able to infer such network models, the behavior of cells and its constituents are experimentally investigated under various experimental conditions. The data is then used to identify the network structure and to estimate the parameters of the model. Usually the output of the model is used to predict the systems behavior and is then experimentally tested and possibly validated, e.g.\ in \cite{becker_covering_2010,bachmann_division_2011}. In addition to point estimates of the model parameters also the uncertainty of the estimates are of interest for the study of the systems. For these nonlinear models, several concepts of assessing the uncertainties and performing identifiability analyses \cite{raue_structural_2009} exist. 

Recently, it has been found for ODE models in a biochemical context, e.g.\ signal transduction or gene regulatory networks, that the eigenvalue spectrum of the Hessian matrix, i.e.\ the second derivative of the objective function with respect to the logarithm of the parameters of almost all examined examples from literature spread over six or even more orders of magnitude \cite{brown_statistical_2003, brown_statistical_2004}. Since this Hessian matrix is closely related to the covariance matrix of estimated parameters, the models have been termed as \textit{sloppy} for the purpose of parameter estimation. Based on this observation, however, it has been claimed, that for ODE models applied in the Systems Biology context this effect is universal \cite{gutenkunst_universally_2007} and it is not possible to reliably estimate parameters \cite{ashyraliyev_gene_2009,bandara_optimal_2009,de_graaf_nutritional_2009}. The same effect has been independently observed in \cite{rand_uncovering_2006,rand_mapping_2008} for circadian clock systems described by ODEs and it has been termed as \textit{rigidity} of the mapping from the parameters to the solutions, where it occupies a space of lower dimension. Moreover, the effect has been shown to emerge naturally in special classes of systems \cite{waterfall_sloppy-model_2006} and a broad eigenvalue spectrum of the Hessian or its analogs is also observed in systems in a general physical context \cite{chachra_structural_2012, machta_parameter_2013}. The impact of the existence of sloppy directions in the parameter space on optimization has been discussed in \cite{transtrum_why_2010, transtrum_geometry_2011}. For ODE models the role of experimental design is well known in general and in the Systems Biology context \cite{balsa-canto_computational_2008,kreutz_systems_2009} and has been shown to be powerful in application settings \cite{raue_identifiability_2010, steiert_experimental_2012}, but \textit{optimal experimental design} has rarely been discussed for sloppy models \cite{apgar_sloppy_2010}.

In this work, we elucidate the origin of the sloppiness effect by analyzing the structure of the covariance matrix. We will illustrate that certain correlations arising from the nature of the used models and experimental setup broaden the eigenvalue spectrum. Knowing the cause of sloppiness in these cases can then be used to show that the sloppy characteristic is crucially influenced by the experimental design and it is possible to find a non-sloppy design for a benchmark model.

\subsection{Parameter estimation}

Biochemical reaction networks can be mathematically described by ordinary differential equations (ODEs), if high copy numbers of the network components can be assumed, so that stochastic effects can be neglected. Furthermore, spatial effects have to be omitted, e.g.\ if diffusion is fast compared to the reaction rates of the network components \cite{wolkenhauer_systems_2008}. Such a system is described by a set of coupled ODEs
\begin{equation}
\dot{\vec{x}}(t) = f(\vec{x}(t),\vec{u}(t), \vec{p}_{dyn})
\label{eq:ode1}
\end{equation}
where in general $f$ is a nonlinear function. $x_i(t)$ denotes the internal states as concentrations of cellular network components at the time $t$ and $u(t)$ denotes time dependent external stimuli or perturbations. Experimentally obtained measurements are represented by 
\begin{equation}
\vec{y}(t)= \vec{g} (\vec{x}(t), \vec{p}_{obs}) + \vec{\epsilon}
\label{eq_ode2}
\end{equation}
where $\vec{g}$ denotes a mapping of one or more internal states $x_i$ contaminated with normally distributed measurement noise $\epsilon \in \mathcal{N}(0,\sigma^2)$. The mapping $\vec{g}$ includes offset and scaling parameters $\vec{p}_{obs}$, whereas Eq.\ (\ref{eq:ode1}) contains the dynamic parameters $\vec{p}_{dyn}$, such as rate constants, Hill coefficients or Michaelis-Menten constants. For model calibration, dynamic and observational parameters, as well as  the initial values $\vec{x}_0$ of the internal states have to be estimated simultaneously in order to find a parameter set $\vec{p}^{\,\ast}=\{ \vec{p}_{dyn}, \vec{p}_{obs}, \vec{x}_0\}$ for which the model describes the experimental data best. Because all parameters are strictly positive and may vary over orders of magnitudes, all analyses is performed on the logarithmic scale.

A common approach of estimating the parameters for an ODE model is solving a non-linear least-squares problem for which the weighted sum of residuals 
\begin{equation}
\chi^2(\vec{p}\,) = \sum^{N_{s,c}}_{j= 1} \sum^{N_{t}}_{i = 1} \left( \frac{y_{ij}^{d} - y_{j}(t_i,\vec{p}\,) }{\sigma_{ij}} \right)^2 
\label{eq:chi2}
\end{equation}
over all observed species $s$, conditions $c$ and time points $t$ has to be minimized where $y_{ij}^{d}$ denotes the measured data and $\sigma_{ij}$ is the measurement error. For Gaussian noise minimizing Eq.\ (\ref{eq:chi2}) is equivalent to a \textit{Maximum Likelihood Estimation} \cite{honerkamp_statistical_2002}. For efficient optimization and to assess the uncertainty of the parameter estimates, the \textit{Sensitivity Equations}
\begin{equation}
\frac{\text{d}}{\text{d} t} \frac{\text{d} x}{\text{d} p} = \frac{\partial f}{\partial x} \frac{\partial x}{\partial p} + \frac{\partial f}{\partial p}
\label{eq_sensitivity_equations}
\end{equation}
are utilzed where the parameter sensitivities 
\begin{equation}
s_{ij} = \left. \frac{\partial x_i(t,u(t),\vec{p}\,)}{{\partial p_j}} \right|_{\vec{p}^{\,\ast}}
\label{sensitivities_single}
\end{equation}
occur as solutions of an ODE which can be attached to the original ODE system and then can simultaneously be integrated \cite{leis_simultaneous_1988}. For least-squares estimation, an approximation of the Hessian matrix
\begin{equation}
H_{\chi^2} \approx  S^{\top} S
\label{eq_hess_sens_approx}
\end{equation}
using the sensitivity matrix $S$ is available under the assumption of additive Gaussian noise for the data \cite{press_numerical_1992}. The sensitivity matrix $S = s_{ij}$ contains the sensitivities of all measurements assigned with index $i$ with respect to all model parameters $p_j$. The size of $S \in \mathbb{R}^{M \times N}$ is given by the number of model parameters $N$ and by the total number of measurement points $M$, which depends on the number of measured observables, the number of applied experimental perturbation and on the chosen time points of recording. The experimental design determines which measurements are performed and determines the structure of the sensitivity matrix $S$ and in turn the structure of the Hessian $H = S^{\top}\, S$ and its eigenvalues. The relationship between the experimental design, the structure of $S$ and the eigenvalues of $H$ is the core topic of the presented work.

\subsection{Uncertainty of parameter estimates}
Typically, in addition to point estimates of the parameters also uncertainties of the estimated parameters are of interest. The covariance matrix
\begin{equation}
C_{\chi^2} \sim \left( H_{{\chi^2}} \right)^{-1}
\label{eq_cov_hess}
\end{equation}
can be used to derive asymptotical confidence intervals. Within such a quadratic approximation of $\chi^2$, contour lines with $\chi^2(\vec{p}\,)=\text{const.}$ are given by ellipsoids in the parameter space. The $\alpha$ quantile $\Delta_{\alpha}(\chi_{df}^2)$ of the  $\chi^2_{df}$ distribution with $df$ degrees of freedom provides borders of the such a confidence ellipsoid, i.e.\ all parameter sets within the ellipsoid
\begin{equation}
E_\alpha = \left\{ \vec{p} \,|\, \chi^2(\vec{p}\,) \leq \chi^2(\vec{p}^{\,\ast}) + \Delta_{\alpha}(\chi_{df}^2)  \right\},
\label{eq_ellipsoid_symp}
\end{equation}
centered around $\vec{p}^{\,\ast}$ are in sufficient agreement with the data. Thus, confidence intervals of the parameter estimates $p_i$ can be obtained by the projection of the ellipsoid $E_\alpha$ onto the $i$-th parameter axis. Typically, such a confidence interval is described by
\begin{equation}
\sigma_i = \sqrt{ \Delta_{\alpha}(\chi_{df}^2) \cdot C_{ii} } \,
\label{eq_se_cov}
\end{equation}
for a parameter $p_i$ where $C_{ii}$ denotes the diagonal entry of the covariance matrix \cite{press_numerical_1992}. For a degree of freedom $df=1$ of the $\chi^2_{df}$ distribution, i.e.\ for point-wise confidence intervals, these $\sigma_i$ are known as the \textit{standard errors}. Applying a $\chi^2_{df}$ distribution with a degree of freedom identical to the number of parameters yields a description for \textit{simultaneous} confidence intervals.
 
Apart from the shortened description of the shape of the iso-$\chi^2$ ellipsoid by the projection on the basis axes of the parameter space, the principal axes of the ellipsoid can be assessed by another analysis of the covariance matrix $C$ of the estimated parameters: The width along the $i$-th principal axis of the ellipsoid is proportional to the square root of the eigenvalue $\lambda^C_i$ of the covariance matrix and the corresponding eigenvector $\vec{e}_{\lambda^C_i}$ points in the direction of the same principal axis. Analogously, one over the square root of the eigenvalues of the Hessian $H$ are likewise proportional to the widths of the ellipsoid along the principal axes. The eigenvector of the smallest eigenvalue of the Hessian matrix points in direction of the largest uncertainty in the parameter space. Since this direction may be a combination of parameters, in general no information about the uncertainty of single parameters is obtained by this description. For a wide spread of the eigenvalues of the Hessian, likewise the principal axes of the confidence ellipsoid differ considerably. The ratio of the eigenvalues, i.e.\ the relative form of the eigenvalue spectrum, does not depend on a certain confidence level. Directions of eigenvectors with small eigenvalues of the Hessian, i.e.\ large widths of the ellipsoid are termed \textit{sloppy}, whereas directions with large eigenvalues are called \textit{stiff}.

\subsection{Sloppiness}

The term \textit{sloppiness} or \textit{sloppy models} in Systems Biology has its origin in \cite{brown_statistical_2003} and \cite{gutenkunst_universally_2007}, where the authors claim sloppiness to be a universal property of the nonlinear models. A quantitative definition of sloppiness is given in \cite{brown_statistical_2003} where the eigenvalue spectrum of the Hessian, with eigenvalues
\begin{equation}
\lambda_i^{norm} =  \frac{\lambda_{i}}{\lambda_{max}}
\end{equation}
normalized by their maximal value, is analyzed. Consequently, the spread of the spectrum is given by the smallest normalized eigenvalue $\lambda_{min}^{norm} =  \lambda_{min}/\lambda_{max}$. Thus, the width on the logarithmic scale of the eigenvalue spectrum is $w^\lambda = \log_{10}(\lambda_{min}^{norm})$. A uniformly distributed eigenvalue spectrum with a width $w^{\lambda} \geq 6$ is referred to a \textit{sloppy} model, whereas a \textit{non-sloppy} model would have $w^{\lambda} \approx 2$, c.f.\ \cite{brown_statistical_2003}. Since a precise discrimination is not given in the known literature, we define non-sloppy models by $w^{\lambda} \leq 3$ for the use of this paper. 

In \cite{gutenkunst_universally_2007}, 17 models from the Systems Biology literature have been analyzed. The normalized eigenvalues of 16 of the examined models individually show a sloppy spectrum: Roughly uniformly distributed eigenvalues on the logarithmic scale over many decades with a minimal normalized eigenvalue $\lambda^{norm}_{min} \approx 10^{-6}$. According to the authors' statement \cite{gutenkunst_universally_2007} this sloppiness is a potential explanation to
\begin{quotation}
``[...] \textit{the difficulty of extracting precise parameter estimates from collective fits, even from comprehensive data.}''
 \end{quotation}
Nowadays, the term \textit{sloppiness} is often found in the literature and sometimes erroneously used for concluding that parameters cannot be estimated and identified and thereby is sometimes utilized as an excuse that the parameter estimation behaves poorly \cite{kirouac_cell-cell_2009,fomekong-nanfack_inferring_2009,cirit_data-driven_2012}.

\section{Origin of Sloppiness \label{chap:cause}}

To understand the origin of sloppiness, the structure of the Hessian matrix $H$ has to be investigated. Using the approximation $H=S^{\top}S$ (cf.\ Eq.\ (\ref{eq_hess_sens_approx})) only the sensitivity matrix $S$ determines the eigenvalue spectrum of the Hessian. The sensitivities $s_{ij}$ (Eq.\ (\ref{sensitivities_single})) quantify the change in the solutions of the ODE system under the change of the estimated parameters and the choice of experiments determines which sensitivities are incorporated into $S$. Thus, the design of the measurements defines the structure of the sensitivity matrix and the Hessian matrix. 
 
In order to learn about the cause of a broadened eigenvalue spectrum of the Hessian, analytical results from \textit{Random Matrix Theory} \cite{edelman_random_2005,mehta_random_2004} are discussed first. Random Matrix Theory has been useful in many areas, such as quantum and nuclear physics \cite{firk_nuclei_2009}, but also in the context of wireless communications \cite{tulino_random_2004} or in the analysis of stock data in the economical science community \cite{plerou_random_2002, laloux_noise_1999}. Starting from uncorrelated entries of the sensitivity matrix, structures as they appear in the considered kind of models and measurements are stepwise incorporated in order to mimic realistic sensitivity matrices and understand the formation of their eigenvalue spectrum. 

\subsection{Marcenkov-Pastur Law \label{sec:MP-law} }

As a starting point, matrices with a minimal level of structure are analyzed. For that purpose randomly distributed entries for the sensitivity matrix are utilized. This artificially mimics an idealized setting in which measurements are independent in terms of their sensitivities. 

For a matrix $X \in \mathbb{R}^{M \times N}$ with i.i.d.\ entries $x_{ij}$, the matrix $W=X{^\top}X$ is commonly known as the \textit{Wishart matrix} \cite{wishart_generalised_1928}. For this natural distribution of eigenvalues of a matrix $W$, an explicit non-random form $f(\lambda)_{MP}$ exists for the limit of $N,M \rightarrow \infty$ which is described by the \textit{Mar\v{c}enkov-Pastur distribution} \cite{marcenko_distribution_1967}.

\begin{figure}[h!]
	\includegraphics[scale=0.44]{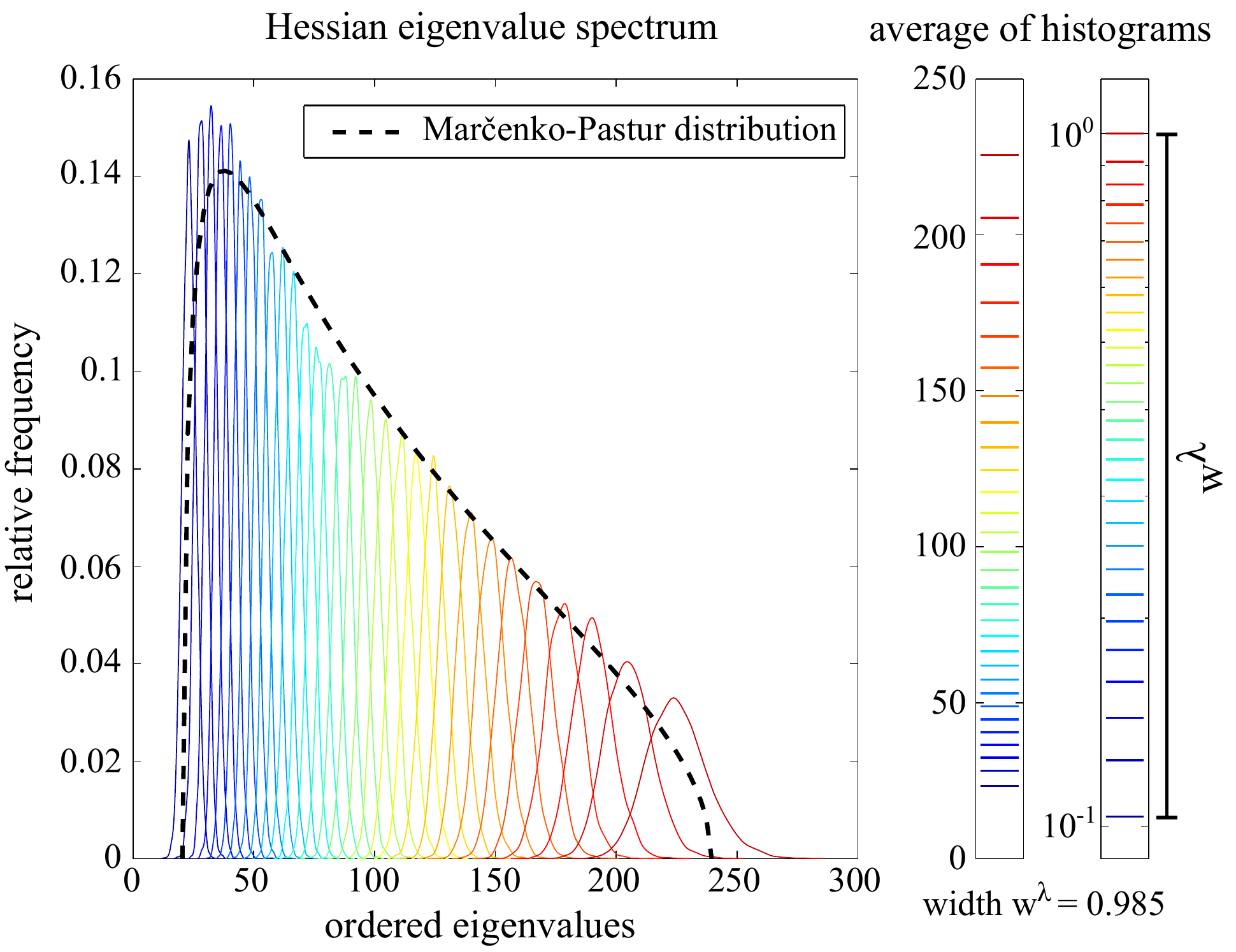}
	\caption{\label{fig:mp_example}(Color online) Histograms of ordered eigenvalues of $H=S^{\top}S$ for $N_S$ = 5000 realizations sensitivity matrices $S$ with $N=30$, $M=100$ and i.i.d.\ entries $s_{ij}$. The corresponding Mar\v{c}enkov-Pastur distribution \cite{marcenko_distribution_1967} is represented by the black dashed line. The averaged eigenvalues show a finite, non-sloppy spectral width.}  
\end{figure}

In our case, the i.i.d.\ entries of the sensitivity matrix $S \in \mathbb{R}^{M \times N}$ are the sensitivities $s_{i,j}$ and the Wishart matrix $H=S{^\top}S$ is identified with the Hessian matrix. Fig.\ \ref{fig:mp_example} shows the histograms of the ordered eigenvalues of $H$ from $N_S= 5000$ realizations of such a matrix $S$ with $N=30$ parameters and $M = 100$ measurement points and the accordance with $f(\lambda)_{MP}$. Although the Mar\v{c}enkov-Pastur distribution is derived for $N,M \rightarrow \infty$, the eigenvalue distribution of the Hessian $H$ from smaller sensitivity matrices $S$ as they appear in applications is in good agreement with the analytical form. Not surprisingly, a sensitivity matrix with i.i.d.\ entries shows a clearly non-sloppy eigenvalue spectrum with width of the averaged eigenvalues $w^{\lambda} = 0.985$. In this setting, neither information about the structure of the system, nor information about the measurements and observables enters. The eigenvalue spectrum has a finite nonzero width $w^{\lambda}$ which is influenced only by the proportion of model parameters to data points. 
 
A more realistic sensitivity matrix would contain certain structures in vertical direction arising from the kind of data which is analyzed and likewise in horizontal direction from the dependencies between the parameters given by the model structure. In the following, these kind of structures are added subsequently to $S$, mimicking the mentioned motives.

\subsection{Integration of time course data structure \label{cause_vert}}

\subsubsection{Autocorrelations in the rows of the sensitivity matrix of a single observable}

In practice, time course experiments are performed in order to assess the dynamics of an ODE model. For a dense time sampling of measurements, the sensitivities of adjacent data points would change almost continuously and therefore are highly correlated. In turn, for larger sampling intervals the correlation decreases. To formalize the vertical correlation in the sensitivity matrix related to such a time course measurement, entries from a first order autoregressive process (AR(1)) can be used to mimic the sensitivities $s_{ij}$ for a general model. Sensitivities from such an AR(1) process are described by 
\begin{equation}
s_{1}^{j} = a_0 + b_0 \epsilon_{1j}  \,, \,\,\,  s_i^{j} = a_1\, s_{i-1}^{j} + b_0  \epsilon_{ij} \,\,\,\, \text{for}\,\,\, i\geq 2
\end{equation}
for the $j$-th parameter at time point $i$ with $\epsilon_{ij} \in \mathcal{N}(0,1)$.  The coefficient $a_1$ controls the correlation between two consecutive time points and $b_0$ is the standard deviation of dynamic noise. This leads to a correlation in vertical direction in the sensitivity matrix
\begin{equation}
S= \left( \begin{array}{cccc}
s_1^{1} & s_1^{2} & ... & s_1^{N} \\
s_2^{1} & s_2^{2} & ... & s_2^{N} \\
\vdots & \vdots & \ddots & \vdots \\
s_M^{1} & s_M^{2} & ... & s_M^{N} \\
\end{array} \right)\,,
\end{equation}
where the sensitivities of a time course measurements with respect to the parameter $p_j$ develop from an independent initial value $s_{1}^{j}$ from the top to the bottom of the matrix in the $j$-th column. Here, identical coefficients $a_0, a_1$ and $b_1$ are chosen for all columns.

\begin{figure}
	\includegraphics[scale=0.33]{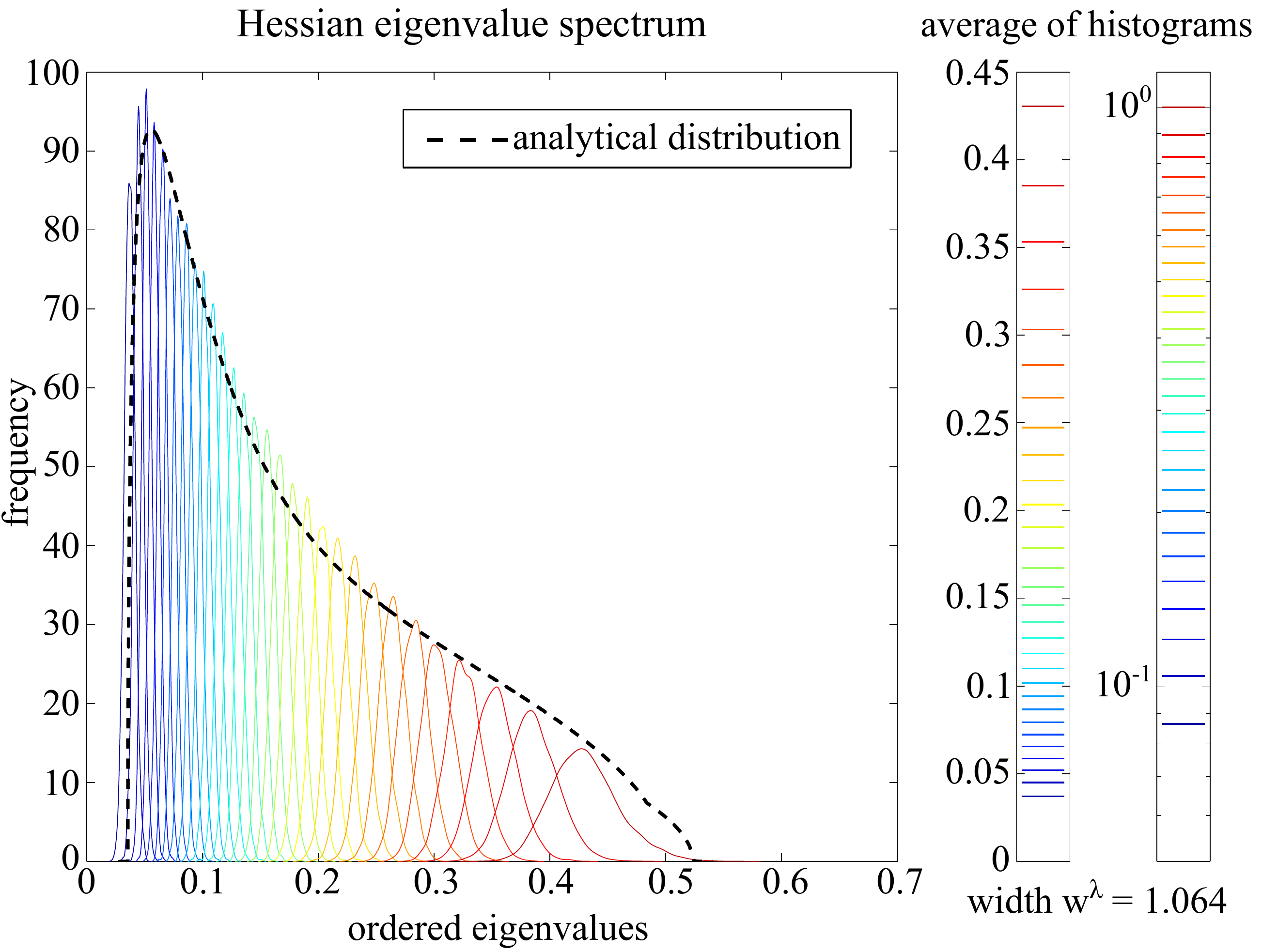}
	\caption{\label{fig:ar1_example}(Color online) Histograms of ordered eigenvalues of the Hessian for $N_S = 5000$ realizations of AR(1) structured sensitivity matrices $S$ with coefficients $a_1 = 0.3, b_0 = 0.4$ for M = 100 measurement points and N = 30 parameters. The black dashed displays the corresponding analytic distribution from \cite{jin_limiting_2009,bai_large_2008} which is in good agreement with the empirical results. The width of the distribution of the averaged eigenvalues on the logarithmic scale $w^\lambda =1.06$ is slightly larger than for uncorrelated entries in the sensitivity matrix.}
\end{figure}

The eigenvalue distribution of a normalized Hessian matrix $H=\frac{1}{M} S^{\top}S$ with such sensitivity matrices $S$ in the limit $M \rightarrow \infty$ has an explicit form $f(\lambda)_{AR(1)}$ {\cite{jin_limiting_2009, bai_large_2008, burda_random_2010}. Likewise to the Mar\v{c}enko-Pastur distribution, the explicit form $f(\lambda)_{AR(1)}$ is derived in the limit of $M \rightarrow \infty$, but shows a good agreement with the empirical eigenvalue distribution for a large number of realizations of samples, cf.\ Fig.\ \ref{fig:ar1_example}. The expected observation occurs that sensitivity matrices with the introduced correlation of the entries yield a slightly wider spectrum on the logarithmic scale compared to sensitivity matrices with uncorrelated entries. Since the explicit form $f(\lambda)_{AR(1)}$ only holds for the limit of $M \rightarrow \infty$ and deviations from the empirical results are observed, e.g.\ for larger values of $a_1$, empirical results, i.e.\ Monte-Carlo simulations will be used to determine the width for other sets of parameters in the following.

\begin{figure}
	\includegraphics[scale=.38]{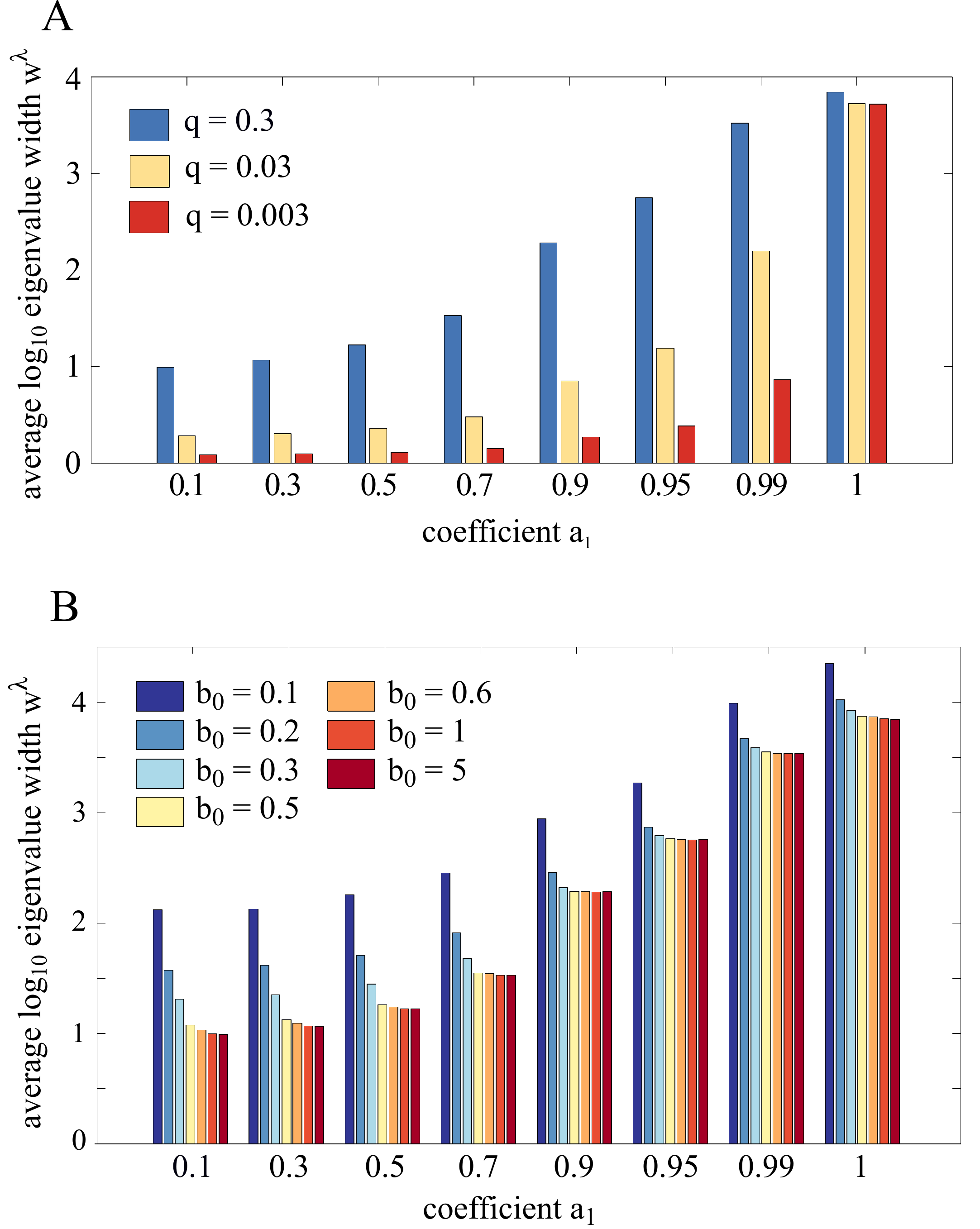}
	\caption{\label{fig:ar_sim_coeffs}(Color online) Width of the eigenvalue spectrum of a Hessian $H=S^\top S$ for sensitivity matrices $S$ with AR(1) structure. The eigenvalues are calculated from 500 samples of $S$ for each parameter set. The spectral widths $w^\lambda$ are averages from the widths of the spectra from the samples. Panel A shows the spectral widths for different coefficients $a_1$ and dimensions ratios $q$ for fixed coefficients $a_0=0, b_0=1$. Panel B shows the spectral widths for sensitivity matrices filled with processes with $a_0 =1$ and their dependency on the coefficients $a_1$ and $ b_0$. The dimensions ratio $q = N/M = 30 / 100$ corresponds to an ODE model with $N=30$ parameters and $M=100$ measurements. The colors of the bars indicate the magnitude of the dynamic noise coefficient $b_0$ in the generating process for the sensitivities. Correlated rows in the sensitivity matrix as obtained for large $a_1$ and small $b_0$ exhibit the largest width of the eigenvalue spectrum.} 
\end{figure}

The shape of the distribution and width of the spectrum of eigenvalues of $H$ is affected by $a_0, a_1$ and $b_0$ as well as by the dimensions ratio $q=N/M$ of the sensitivity matrix. In the setting from which the explicit form $f(\lambda)_{AR(1)}$ is derived with $a_0 = 0$, the coefficient $b_0$ has no influence on the width of the \textit{logarithmic} scale. Large correlations of the sensitivities generated by a process with $a_1$ close to $1$ yields an almost sloppy spectrum of the Hessian with $w^{\lambda} \approx 4$, as displayed in Fig.\ \ref{fig:ar_sim_coeffs}\,A. The dimensions ratio $q = N/M$ corresponds to the number of measurement points $M$, i.e.\ rows in $S$, so that smaller values of $q$ yield a narrower spectrum.

For a generating process with initial value $a_0 = 1$ the dynamic noise coefficient $b_0$ becomes relevant, cf.\ Fig.\ \ref{fig:ar_sim_coeffs}\,B. Again, large values of $a_1$ produce a highly broadened spectrum, up to spectral widths $w^{\lambda} = 4.3$. The combination of large $a_1$ and small $b_0$ yields the highest correlations of adjacent rows in the sensitivity matrix and corresponds to sensitivities of densely sampled measurements.

In a formal mathematical sense, $a_1 \geq 1$ yields an instationary process, i.e.\ the variance of the process is increasing with time. However, since the measurement times in applications constitute finite time intervals, violating stationarity is not a serious issue here. In turn, since the autoregressive process is used to imitate a deterministic process, choosing $a_1 = 1$ and $b_0 \ll 1$ realistically resembles the relationship between the sensitivities of adjacent measurement times. It should be noted that in the literature concerning the sloppiness issue \cite{gutenkunst_universally_2007}, $\chi^2$ was integrated along the time axis which corresponds to a densely time point sampling and thus may create such highly correlated sensitivity time courses. In terms of optimal experimental design, this is a setting enhancing sloppiness. In turn, narrow spectral widths are achieved if a small coefficient $a_1$ is assumed which implies that the correlation of the sensitivities of consecutive time points is small, i.e.\ the dynamics of the observed signal is fast compared to the temporal measurement point sampling. 

Summing up, the simulation study presented in this section shows that for strongly correlated rows of the sensitivity matrix, as obtained for a dense time sampling of a single observable, e.g.\ for a dimensions ratio $q = 0.03$ causes a more than 10-fold broader eigenvalue spectrum than the natural Mar\v{c}enko-Pastur distribution. Moreover, the findings guide to possibilities for optimal experimental design methods reducing the sloppiness, e.g.\ by choosing only several representative measurement time points.

\subsubsection{Multiple experiments cause block structure \label{chap:multiple_exps}}

Typically, measurements from more than one observable are needed to gain enough information about the system in order to estimate the parameters of an ODE model. Furthermore, perturbations on reactions in the system can be performed experimentally in order to get additional information about the system. Thus, a further step in mimicking the structure of a realistic sensitivity matrix is the incorporation of several observables or data obtained for perturbed settings, which results in a block-wise structure of the sensitivity matrix $S$. 

Within a block of a measurement, the sensitivities are simulated by a process accordingly to the previous analysis of the AR(1) processes with equivalent coefficients $a_1$ and $b_0$. The independency between the blocks corresponds to different initial values $s_1^{j} \in \mathcal{N}(0,1)$ for each time course, i.e.\ block of the sensitivities in horizontal and vertical direction, in order to mimic the most general case of a model structure and choice of experiments. In our simulation the same length is chosen for all blocks in a sensitivity matrix, i.e.\ the same number of data points per experiment. The total number of measurements is kept constant to $M = 1000$ and the number of parameters is $N=30$ for all examined matrices in the further procedure.

\begin{figure*}[!htbp]
	\includegraphics[scale=.5]{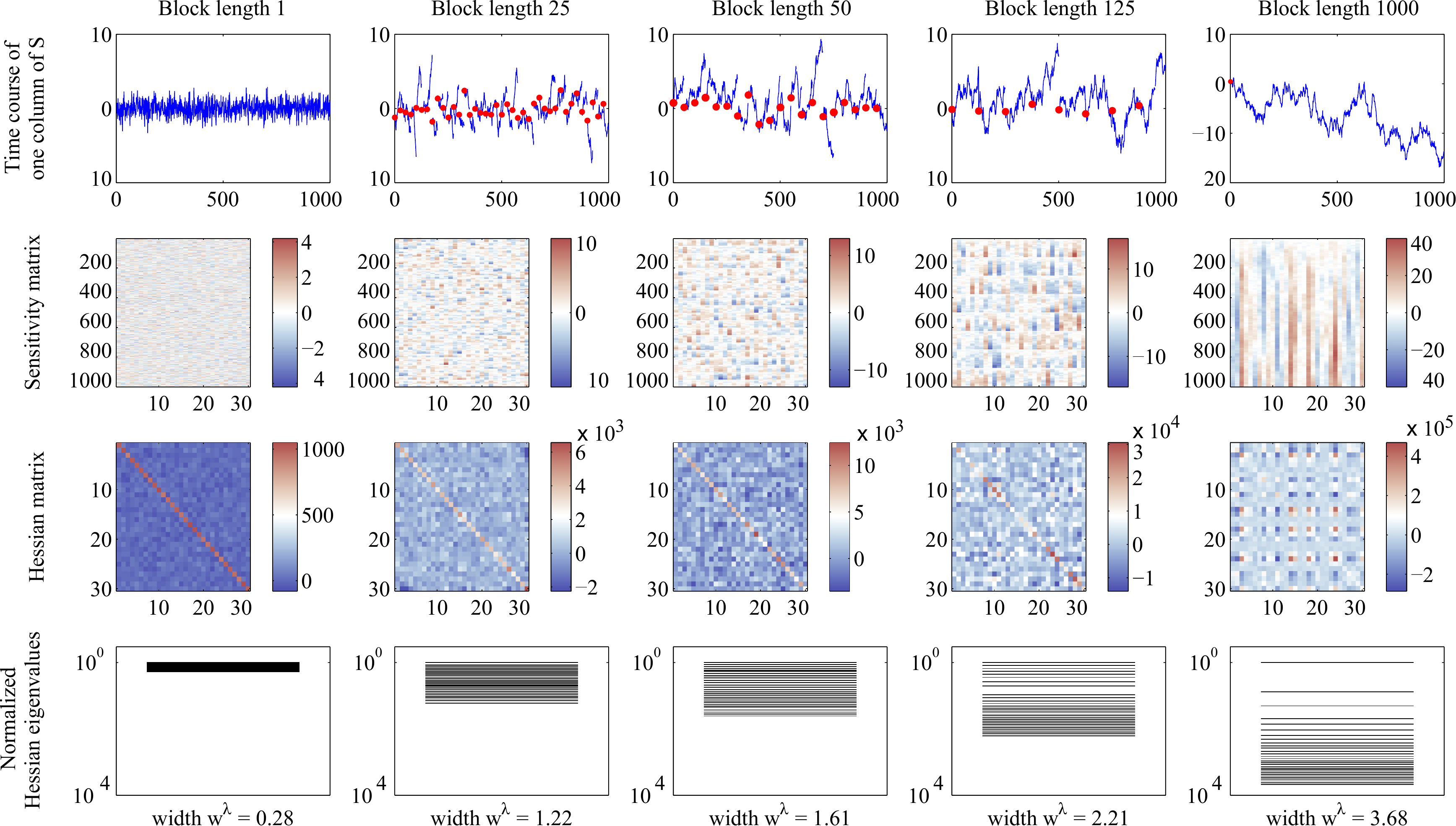}
	\caption{\label{fig:blocks_ninfo}(Color online) Eigenvalue spectra for independent blocks in the sensitivity matrix for different block lengths. The sensitivities within a block are generated by an autocorrelated process with coefficients $a_1 = 1$ and $b_0 = 0.5$ and randomly chosen initial values for each block $s^{j}_1 \in \mathcal{N}(0,v^2_0)$, with standard deviation $v_0 = 1$. The upper row shows an exemplary time course of one column of $S$ where red dots indicate the start of a new independent block. The second and third row show the sensitivity matrix and the Hessian, respectively. The lower row illustrates the average normalized eigenvalue spectrum of the Hessian with the value of $w^\lambda$ below.  The width of the eigenvalue spectrum with uniformly distributed eigenvalues gradually scales with the length of independent blocks.}  
\end{figure*}

Fig.\ \ref{fig:blocks_ninfo} illustrates such a setting of multiple measurements, the structure of the resulting sensitivity matrix $S$ and Hessian $H = S^\top S$, as well as the corresponding eigenvalue spectrum of $H$ for different block lengths. Tuning the block length $L$, the eigenvalue spectrum of $H$ can be arbitrarily scaled between the Mar\v{c}enko-Pastur case corresponding to a block length of one, up to a single block with the maximal block length $M$ in the sensitivity matrix which represents the case of an AR(1) structured matrix, as discussed in the previous section. For every analyzed block length between these two cases, the Hessian exhibits roughly equidistant eigenvalues. 

A more sloppy design with $w^\lambda \approx 6$ is obtained by using a slightly different process to generate a structure of multiple plateaus for the sensitivities. For that purpose the standard deviation $v_0$ of the initial sensitivities $s^{j}_1 \in \mathcal{N}(0,v^2_0)$ is increased to $v_0 = 20$. Such a process with coefficients $a_1 = 1$ and $b_0 = 0.02$ generates a structure of multiple plateaus for the sensitivities with varying initial values $s^{j}_{1} = \pm 75$ and fluctuating values of $s_{ij}$ around a constant level for each block within the plateaus, as shown in Fig.\ \ref{fig:blocks_ninfo_split}. Under these constraints, the widths of the eigenvalue spectrum ascend stepwise with an increasing block length up to a block size of $L = 25$ and show nearly uniformly distributed eigenvalues. For the next feasible block length $L=40$, the eigenvalue spectrum increases abruptly to $w^\lambda = 5.99$ and looses its evenly spaced character (cf.\ Fig.\,\ref{fig:blocks_ninfo_split} bottom left). The eigenvalues of the spectrum split up in two groups, from which the upper subgroup contains the same number of eigenvalues as the number of blocks in the sensitivity matrix. In the same way, it is possible to tune the width of the eigenvalue spectrum up to at least ten orders of magnitude by choosing extreme values of $v_0$, $b_0$, and $q$. In such a case, however, the entries of the sensitivity matrices would look less realistic if compared to a typical ODE application setting.

\begin{figure*}[htbp]
	\includegraphics[scale=.44]{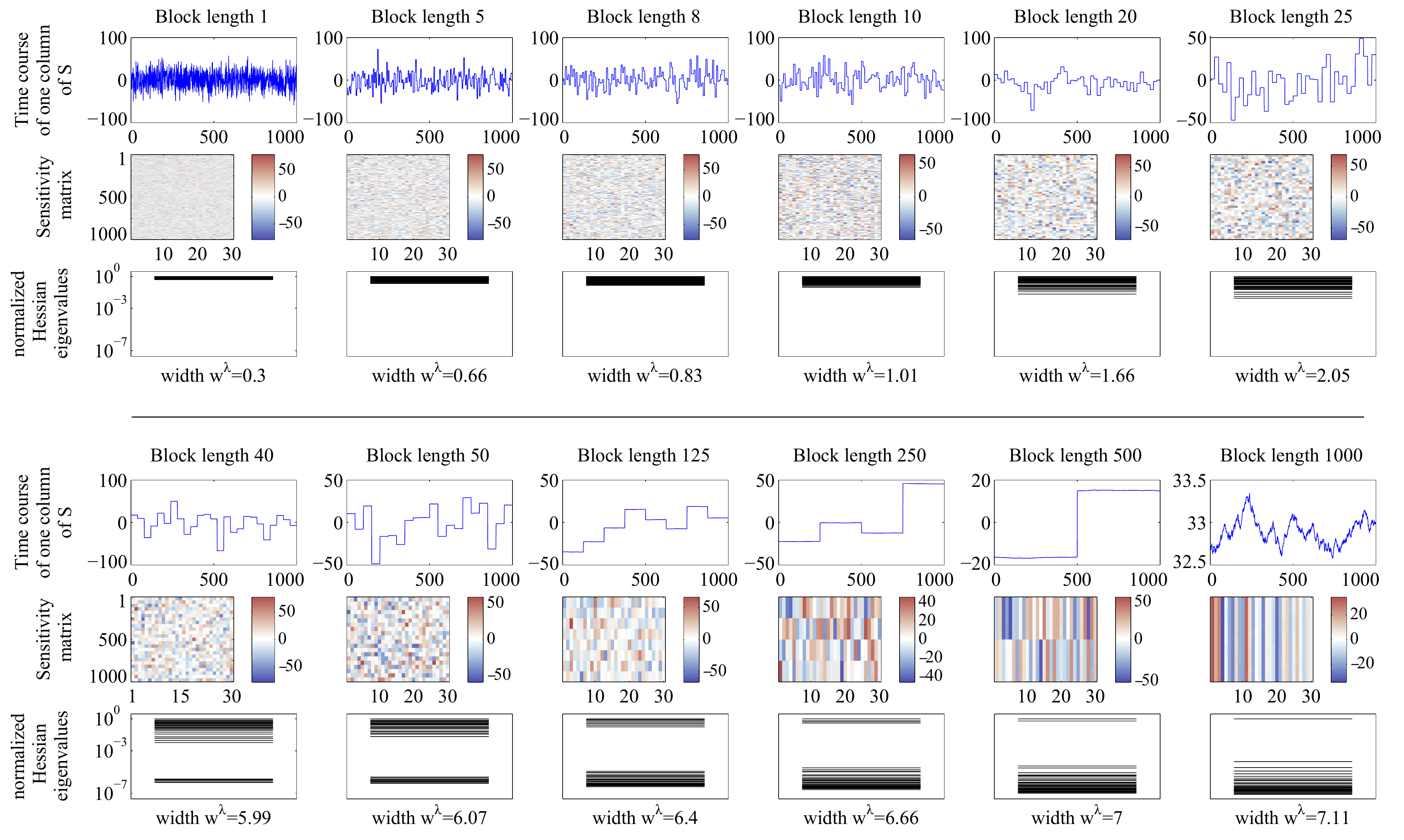}
	\caption{\label{fig:blocks_ninfo_split}(Color online) Sensitivity matrices with independent blocks with coefficients $a_1 = 1, b_0 = 0.02$ and standard deviation of the initial values $v_0 = 20$ of the generating process of the sensitivities, for different block lengths likewise to Fig.\ \ref{fig:blocks_ninfo}. The block length $L$ is changed from left to right in each row. Matrices with $L \geq 40$ have a sloppy spectral width, but the eigenvalues are not uniformly distributed on the logarithmic scale.}  
\end{figure*}

\subsection{Incorporating the model topology}

So far, all parameters were assumed to contribute with a similar impact to the structure of the sensitivity matrix. For example, the randomly chosen initial values of the sensitivities in adjacent blocks in horizontal direction had the same variance and by using the same coefficients of the generating processes, also the variance of the sensitivity time courses is the same in all blocks. However, due to the diverse functions of the parameters, variations of single parameters may yield a considerably different impact on specific observables. Thus, a structure in the sensitivity matrix in horizontal direction emerges. In the following section, this aspect is integrated successively in the simulation of realistic sensitivity matrices showing a sloppy behavior.

\subsubsection{Partitioned sensitivity matrix \label{sec:part_sens}}

Usually, ODE models contain parameters with different physical units. Typically, there are dimensionless parameters like Hill coefficients, rates with dimension one over time, as well as bimolecular reaction rates with dimension one over time and concentration or parameters with the dimension of the concentration unit, e.g.\ Michaelis-Menten constants. They may additionally differ in their biologically reasonable range. Furthermore, the used measurement techniques, e.g.\ for protein concentrations are typically not determined to an absolute value but are recorded in arbitrary units and thus also the units of the model parameters are often not specified. Theoretical results for the distribution of confidence interval sizes or eigenvalues only hold for a group of parameters with the same physical units. Therefore each group of parameters with the same physical unit has to be considered independently. Since variances are additive, the freedom of choosing the parameters in a different unit system, in general, broadens the range of confidence interval sizes as well as the range of observed eigenvalues.

\begin{figure*}[htbp!]
	\includegraphics[scale=.6]{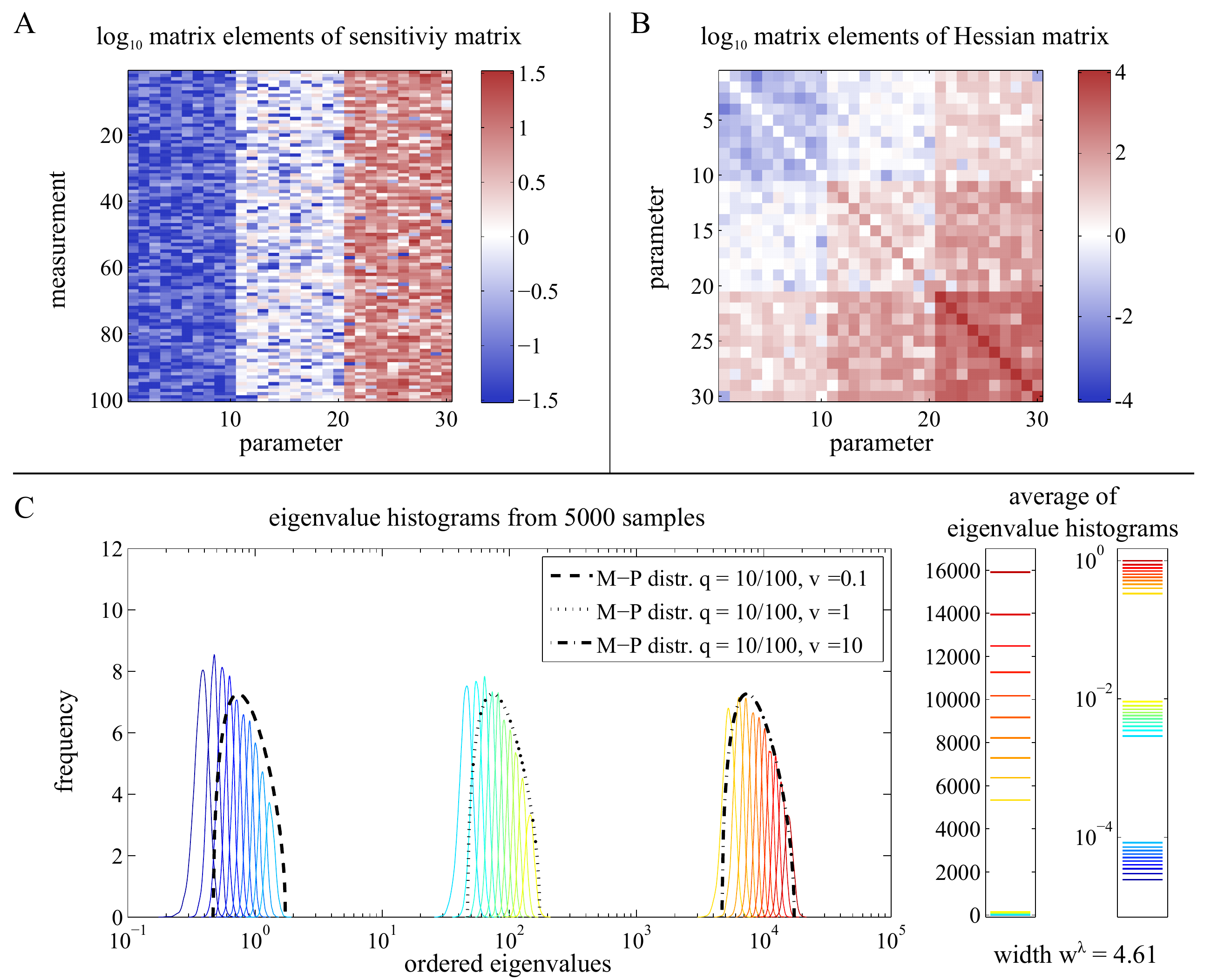}
	\caption{\label{fig:part_sens}(Color online) Panel A shows the partitioned sensitivity matrix with three groups of parameters. To mimic the effect of different physical units within a group of parameters, different standard deviations $v_1\left(p_{1-10}\right)=0.1$, $v_2\left(p_{11-20}\right)=1$, $v_3\left(p_{21-30}\right)=10$ are used. Panel B shows how the structure in the sensitivities translate to the Hessian matrix. For better illustration of the block structure in the matrices the logarithms of the absolute values of the matrix elements are shown in panel A and B. Within each block in the Hessian, the eigenvalue distributions are approximately given by the Mar\v{c}enkov-Pastur distribution (panel C). The difference between the blocks yields three groups of eigenvalues with different location on the logarithmic scale.}
\end{figure*}

Such a grouping of ranges of parameter, depending on their unit or function within the model, also translates into the sensitivities. The sensitivities of parameters belonging to certain groups, can be mimicked for a general case by the variation of the magnitude of the variance $v^2_{\alpha}$ of i.i.d.\ distributed entries $s_{ij} \in \mathcal{N}(0,v^2_{\alpha})$. This yields a \textit{partitioned sensitivity matrix} with a block-wise structure of differently scaled entries. Such a matrix is displayed in Fig.\ \ref{fig:part_sens}\,A for three groups of parameters with different variances $v^2_{\alpha}$ for each group. This structure in the sensitivity matrix $S$ translates to a characteristic structure in the Hessian matrix (Fig.\ \ref{fig:part_sens}\,B) and thus results in a fragmentation of the eigenvalue spectrum. The subgroups have a similar shape like the eigenvalue spectra described by the Mar\v{c}enko-Pastur distribution, as illustrated in Fig.\ \ref{fig:part_sens}\,C. Due to the different variances of the entries between the blocks, also the histograms of the ordered eigenvalues form separated subgroups accordingly to the blocks in the sensitivity matrix.

Interestingly, the shape and the position of the subgroups is roughly captured by the adapted Mar\v{c}enko-Pastur distribution for an altered ratio $q$ with the number of parameters per block $q_p = N_p/M$ and corresponding variance $v_\alpha^2$ of the block, as displayed by the black dashed lines in Fig. \ref{fig:part_sens}\,C. The width within the resulting subgroups in the eigenvalue distribution on the logarithmic scale is roughly the same for all three groups. The distance between the subgroups in the eigenvalue spectrum varies with the variance $v^2_{\alpha}$ of the entries of the sensitivity matrix, whereas the individual form of the subgroups in the spectrum is determined by the adapted ratio $q_p = N_p/M$, where $N_p$ is the number of parameters in the corresponding block in $S$. Hence, the width of the whole eigenvalue spectrum can be broadened arbitrarily by increasing the value $\Delta v^2 = v^2_\alpha - v^2_{\alpha-1}$ by which the variances of the entries vary between the subgroups. Larger values of the variance difference $\Delta v^2$ yield a broadened spectrum which can also reach the sloppy setting with $w^\lambda \ge 6$. This kind of structure in the sensitivity matrix constitutes very likely an additional reason for the sloppy shape of the eigenvalue spectra of the literature models. 

\subsubsection{Sparsity and block structure mimic sloppy real world case}

After the simulation analyses of general structures in the sensitivity matrix originating from data of densely sampled time course experiments and differing ranges of parameter values, the model structure should be integrated realistically in the analysis of random matrices. But if no specific ODE model is assumed, no concrete horizontal structure can be integrated without restricting the analysis to a specific model. However, regularities in horizontal direction are obvious which stem from the model topology: Many measurements are sensitive only for a specific set of parameters where the sensitivities change significantly within one time course. Apart from that, the sensitivities for all other parameters are almost or even exactly zero within an experiment. This signifies that some measurements are completely insensitive to certain parameters. For example, a measurement of a species upstream in the system will not be sensitive to a parameter which controls a downstream reaction and thus does not influence the measured observable directly or indirectly.
 
 \begin{figure*}[htbp!]
	\includegraphics[scale=.5]{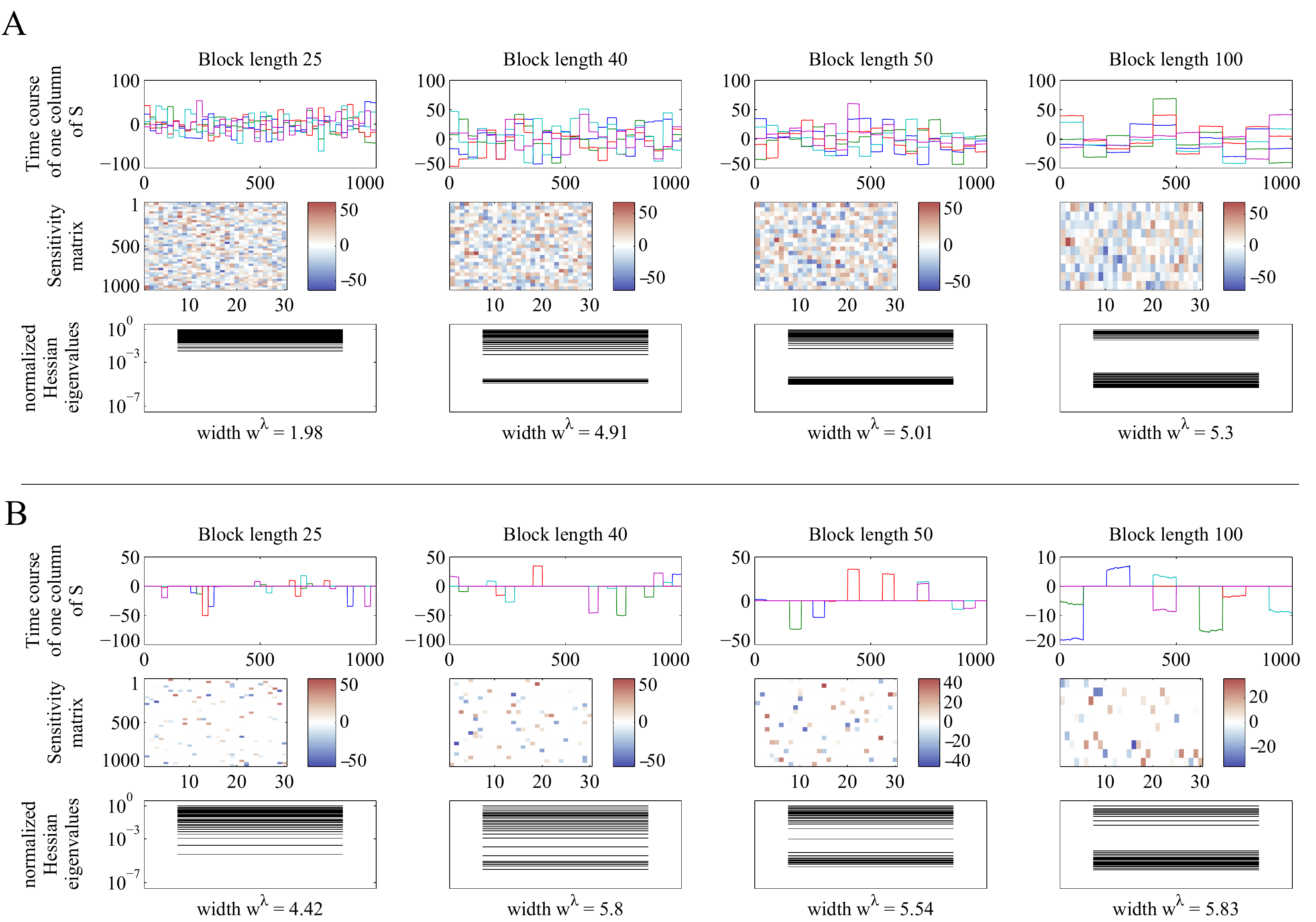}
	\caption{\label{fig:blocks_sparse}(Color online) The upper row in panel A shows the sensitivity time courses of five columns of an exemplary sensitivity matrix $S$, the row in the middle presents the sensitivity matrix $S$ and the lower row illustrates the average eigenvalue spectrum of corresponding Hessian matrices. The entries of $S$ are generated from processes with coefficients $a_1 = 1, b_0 = 0.07$ and $ v_0 =  20$ for four exemplary block lengths. As a reference, panel B shows the outcomes for incompletely filled matrices with only 5\% non zero elements yielding an increased width of the eigenvalue spectrum, similar as for sloppy matrices.}  
\end{figure*}
 
The sparsity effect in the sensitivity matrix is included in our simulations by filling only a certain percentage of the sensitivity matrix with non-zero entries. These nonzero entries are generated by block-wise independent processes as described earlier in Sec.\ \ref{chap:multiple_exps}. Depending on the applied coefficients of the generating process of the sensitivities, the resulting eigenvalues are sloppy. Fig.\ \ref{fig:blocks_sparse} illustrates the difference between 100\% filled sensitivity matrices in the contrast to 5\% filled matrices. The total spread of the eigenvalues increases slightly after removing 95\% of the entries, whereas the gap between the two groups of eigenvalues almost vanishes for a medium block length $L = 40$. 

The simulation results show a sloppy behavior of the resulting eigenvalue spectrum with a spectral width $w^\lambda \approx 6$ and almost uniformly distributed eigenvalues similar to the spectra of the investigated models from \cite{gutenkunst_universally_2007}. Hence, sensitivity matrices randomly filled with blocks containing highly correlated entries in vertical direction are able to explain a sloppy eigenvalue spectrum of the Hessian, although no specific information about a certain model or measurement was introduced.

\begin{figure}[htbp]
	\includegraphics[scale=.42]{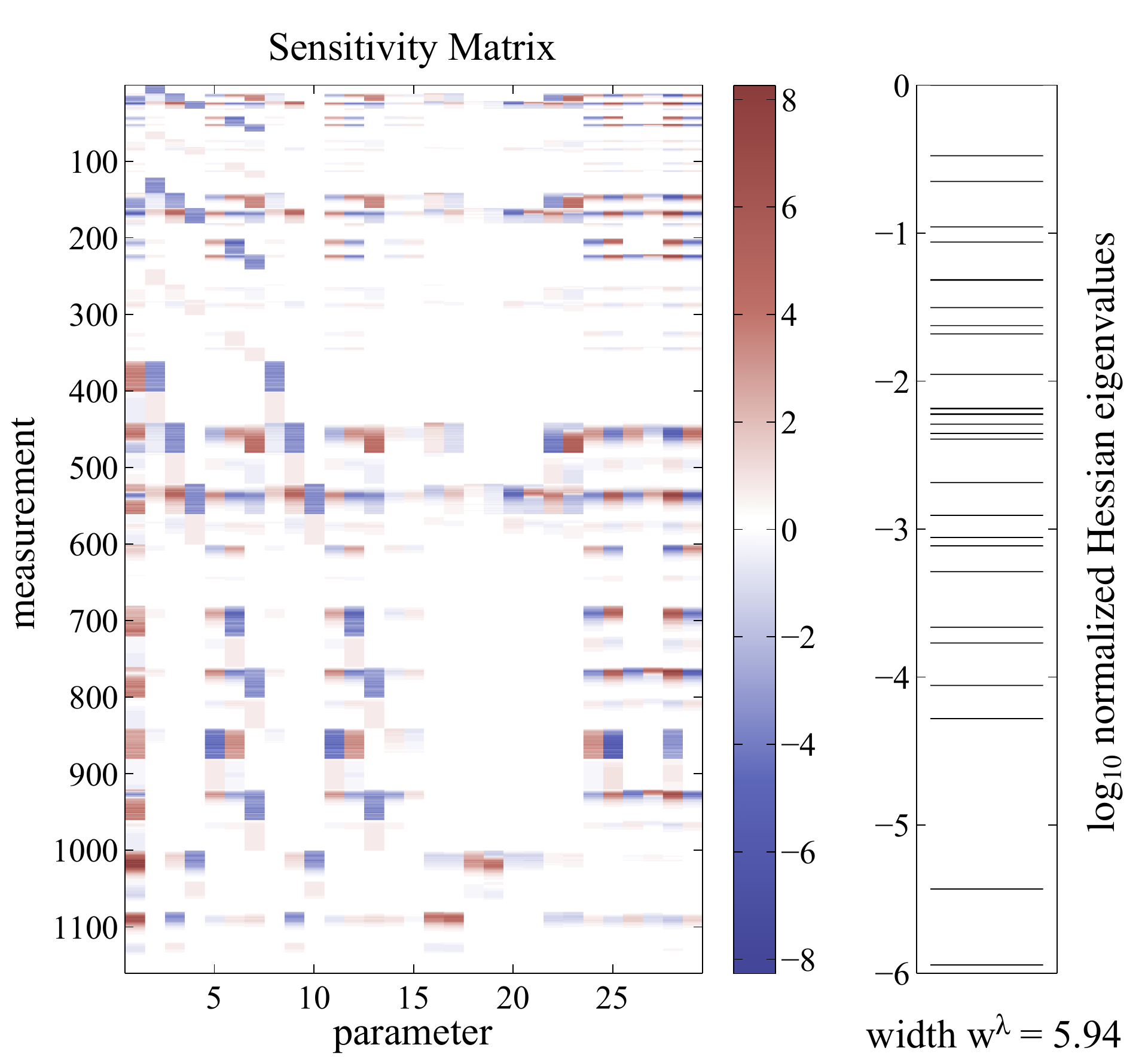}
	\caption{\label{fig:realworld_sens}(Color online) Exemplary real world sensitivity matrix from a typical intermediate step of the experimental design procedure of the DREAM 6 Parameter Estimation Challenge \cite{dream6_paraest}. The colors correspond to the square root of the absolute values of the entries of the sensitivity matrix ${s}^{plot}_{ij}=sgn(s_{ij})\cdot \sqrt{|s_{ij}|}$ which provides the best insight about the structure occurring in the matrix.}  
\end{figure}

In Fig.\ \ref{fig:realworld_sens} a sensitivity matrix of a real world sloppy example from model 1 from the DREAM 6 Parameter Estimation Challenge \cite{dream6_paraest}, which will be introduced as a benchmark model later in Sec.\ \ref{chap:benchmark}, is depicted. The comparison of such sensitivity matrices of time course measurements from applications with the simulated sensitivity matrices shows a good agreement of the discussed block structure from multiple experiments and the insensitivity to the bulk of parameters of most measurements.

\subsection{Summary}

\begin{figure}[htbp]
	\includegraphics[scale=.35]{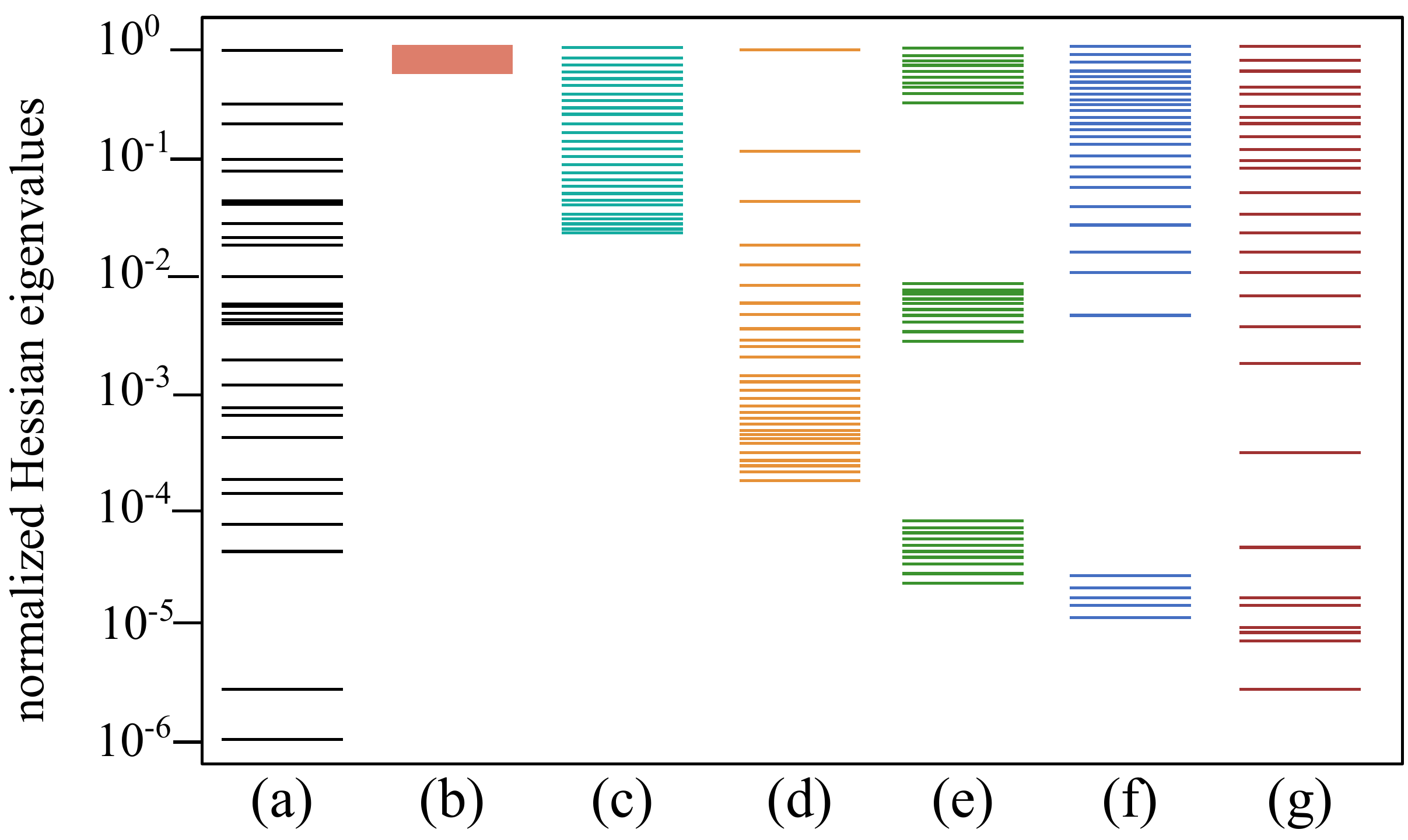}
	\caption{\label{fig:comparison_spectra}(Color online) Comparison of normalized eigenvalue spectra of the Hessian matrix: The example from an intermediate experimental design stage of model 1 from the DREAM 6 Parameter Estimation Challenge \cite{dream6_paraest} with 29 parameters and $1160$ data points shows a sloppy spectrum (a). Averages of eigenvalue spectra from simulations of mimicking the sensitivity matrix with 30 parameters and 1000 data points: (b) matrix randomly filled with i.i.d. entries (Mar\v{c}enko-Pastur distribution), (c) block structured matrix mimicking multiple experiments with block length $L = 50$, (d) matrix filled with a single AR(1) block of the sensitivities with coefficients $a_1=1, b_0=1$, (e) partitioned sensitivity matrix, (f) block structured matrix with large initial values with standard deviation $v_0 = 20$ and block length $L = 40$ (g) sparsely filled matrix with block structure with 5\% nonzero entries and block length $L = 40$. The eigenvalue spectrum gradually increases with the more defined structure in the sensitivity matrix up to a setting with a typical sloppy case with $w^\lambda \approx 6$ and logarithmically uniformly distributed eigenvalues.}  
\end{figure}

The sloppiness effect could be reproduced for a general setting of an ODE model and a general experimental setup of time course data mimicked by entries of the sensitivity matrix with a certain correlation structure. Starting from unstructured, non-sloppy eigenvalue spectra of random matrices, the introduction of correlations corresponding to structures from densely sampled time course data and multiple observables already gradually broadens the spectral width of the Hessian over the non-sloppy limit of $w^\lambda = 3$, as illustrated in Fig.\ \ref{fig:comparison_spectra}. Inserting additional structures arising from the properties of the model and the insensitivity of some parameters on certain measurements further increased sloppiness characteristics. In general, sensitivity matrices which are sparsely filled with blocks of highly correlated entries yield eigenvalue spectra of the Hessian (cf.\ (g) in Fig.\ \ref{fig:comparison_spectra}) comparable to a sloppy real world example (cf.\ (a) in Fig.\ \ref{fig:comparison_spectra}).

\bigskip

\section{Cure of Sloppiness \label{chap:cure}}

As presented in the previous chapter, structures in the sensitivity matrix influence the eigenvalue spectrum of the Hessian matrix. In order to control the width of the spectrum and to circumvent the sloppiness effect for a benchmark model, optimal experimental design methods are applied which directly target on the structure of the sensitivity matrix.

\subsection{Benchmark model and experimental setup \label{chap:benchmark}}

The following analysis will be performed on the basis of a benchmark model from the \textit{DREAM 6 Parameter Estimation Challenge} \cite{dream6_paraest}, for which three artificial, biologically motivated \textit{gene regulatory networks} (GRN) were investigated. In this challenge, a set of possible experiments was defined and a limited number of simulated measurements could be virtually purchased in order to gain information about the model and to estimate its parameters. For a virtual measurement, the observable, the experimental perturbation and the time sampling had to be specified to generate simulated data using the true model parameters for the ODEs. Noise was added to the data to mimic the measurement error. The task of the challenge was to stepwise select informative experiments and to use the data for the parameter estimation and uncertainty analysis which was judged by a certain score.

Fig.\ \ref{fig:dream_model_structure} shows the used model structure with 29 dynamic parameters. The dynamics of the species $x_i$ of the model are given by first-order ODEs. Each model equation is implemented by a sum of fluxes of the respective species. A regulatory reaction v$_j$ is modeled by Hill kinetics and is represented by two parameters, a Hill coefficient $p_{Hill_j}$ and a Michaelis-Menten constant $p_{Kd_j}$. Hence, e.g.\ the transcription of the gene for protein 4 which is activated by protein 1 and repressed by protein 5 has the rate
\begin{widetext}
\begin{equation}
\dot{x}_{mRNA4} = p_{syn, mRNA4} \cdot \underbrace{ \frac{ \left( \frac{x_{pro1}}{p_{Kd_1}} \right)^{p_{Hill_1}} }{ 1 + \left( \frac{x_{pro1}}{p_{Kd_1}} \right)^{p_{Hill_1}} } }_{\text{Activation}} \cdot  \underbrace{\frac{ 1 }{ 1 + \left( \frac{x_{pro5}}{p_{Kd_8}} \right)^{p_{Hill_8}} } }_{\text{Inhibition}} \,-\, \,p_{deg, mRNA4} \cdot x_{mRNA4}\,.
\label{eq:dream_pp4}
\end{equation}
\end{widetext}
The parameter $p_{syn, mRNA4}$ refers to the promoter strength and regulates the mRNA synthesis, whereas the last term of Eq.\ (\ref{eq:dream_pp4}) is responsible for the degradation of the mRNA, implemented by a linear decay with the parameter $p_{deg, mRNA4}$. The translation of mRNA into a protein is given by a linear ODE, e.g.\ for protein 4 it holds
\begin{equation}
\dot{x}_{pro\,4} = p_{syn, pro4} \cdot x_{mRNA\,4}  \,-\, \,p_{deg, pro4} \cdot x_{pro\,4}\,,
\label{eq_dream_p4}
\end{equation}
where the first term determines the synthesis of protein $x_{pro4}$, controlled by the parameter $p_{syn, pro4}$, which is interpreted as the strength of the ribosomal activity site and the degradation of the protein $x_{pro\,4}$ is described by the parameter $p_{deg, pro4}$. Furthermore, as initial concentrations of the proteins $x(0)_{pro\,i} = 1$ and for the mRNAs $x(0)_{mRNA\,i} = 0$ are assumed.

\begin{figure}[htbp]
	\includegraphics[scale=.35]{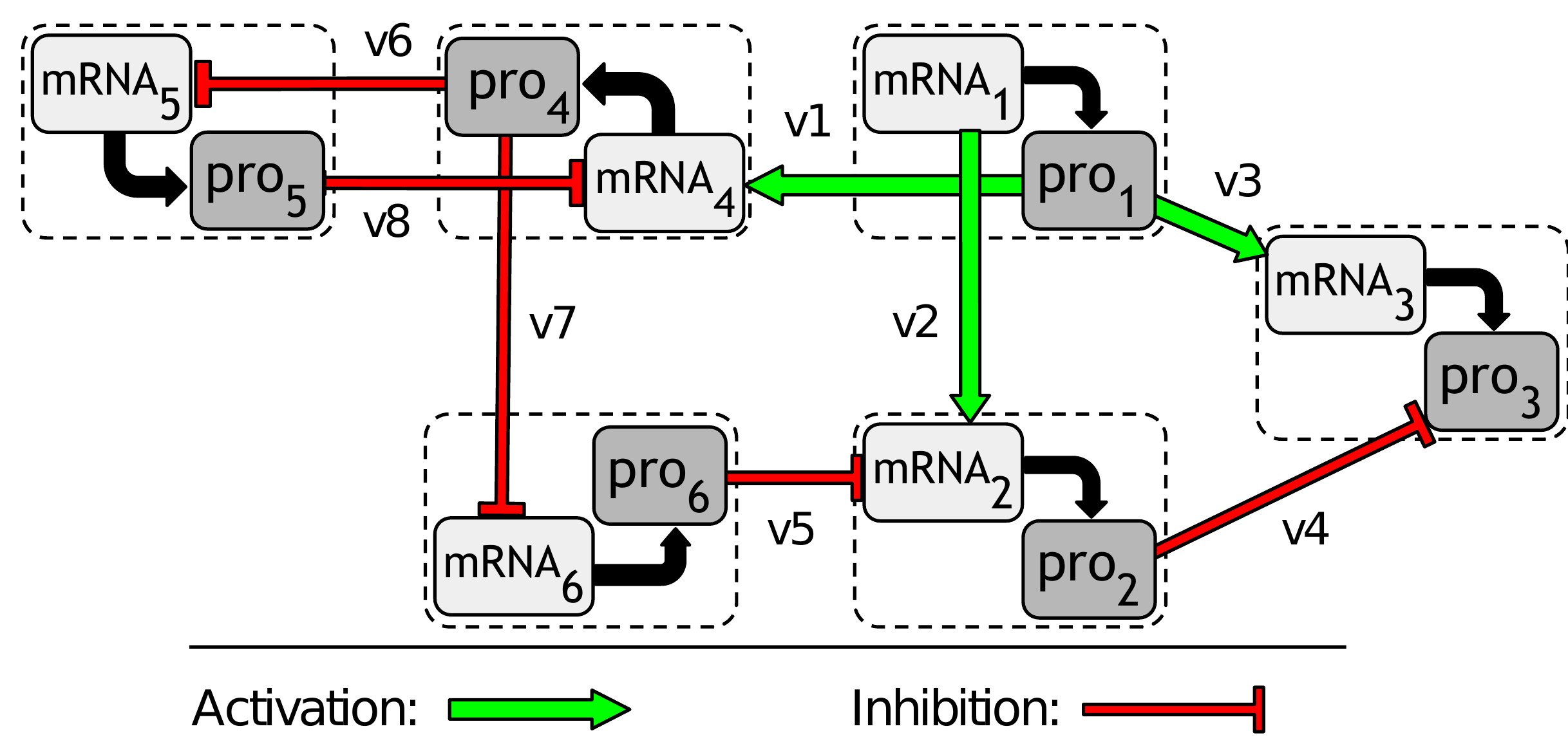}
	\caption{\label{fig:dream_model_structure}(Color online) Model 1 of the DREAM 6 Parameter Estimation Challenge \cite{dream6_paraest}: The network consists of 12 internal states, 6 proteins, denoted by pro$_i$, 6 corresponding messenger RNAs, denoted by mRNA$_i$ and eight regulatory reactions, denoted by v$_i$. Each protein production is implemented by a two step process of transcription from DNA to mRNA and translation of this transcript into a protein. All proteins, except protein pro$_3$, act as a transcription factor which regulates the transcription of other genes in the network.}
\end{figure}

The following analysis of this benchmark model is not only restricted to GRNs, but can also be adapted to other ODE models, e.g.\ to cellular signaling pathway models. For the implementation of the model and parameter estimation, the \textit{D2D software} \cite{raue_lessons_2013} was utilized.

\subsection{Sensitivity matrix structure designs}

The previous simulations in Chap.\ \ref{chap:cause} showed the strong impact of data from time course experiments on sloppiness. Thus, removing the block-wise structure of correlated entries $s_{ij}$ in the sensitivity matrix is a promising approach, as matrices with uncorrelated entries have been shown to create much lower values of $w^\lambda$. Consequently, the task for the experimental planing if sloppiness is intended to be minimized, is to diminish this kind of correlation by selecting sampling times for time course measurements where consecutive data points do not provide redundant information. Since analytic solutions of the sensitivities are available only for special cases, the selection of measurements according to this demand is a non trivial task. 

The horizontal structure in $S$ is determined by the model structure, whereas the vertical structure in the sensitivity matrix results from the choice of experiments. While correlations in vertical direction in $S$ are manageable, e.g.\ by the choice of the sampling times, horizontal structures cannot be controlled sufficiently in the same way because of the given model structure. This hinders the construction of a sensitivity matrix with completely uncorrelated and independent entries, which would lead to a maximally narrow eigenvalue spectrum similar as shown for the eigenvalue spectra of matrices described by the Mar\v{c}enko-Pastur distribution. A possibility to influence the sensitivities in this manner also in their horizontal structure are perturbation experiments, i.e.\ experimentally controlled alterations of parts of reactions in the system, which is an established approach in molecular biology. Perturbations on several targets of the system may be also applied simultaneously to increase the variety of possible structures in $S$ and the flexibility for the experimental planing. The task for the optimal experimental design method is to select these experiments which induce a certain structure in $S$ which then yield to a low value of $w^\lambda$ with a minimal number of experiments in terms of perturbations. 

\begingroup
\squeezetable
\begin{table*}[htbp]
 \begin{ruledtabular}
\begin{tabular}{c  c  c}
 \footnotesize
  perturbation & description  & implementation \\
  \hline 
  gene $i$ knockout & knockout of gene $i$ production & set $p_{\text{syn,pro}i}$ and $p_{\text{syn,mRNA}i}$ to 0 \\
  rbs $i$ over & 2-fold over expression of protein $i$ & doubling of $p_{\text{syn,mRNA}i}$ \\
  rbs $i$ 10over & 10-fold over expression of protein $i$ &  increase of $p_{\text{syn,mRNA}i}$ by factor 10 \\
  siRNA $i$ down & 5-fold down regulation of gene $i$ & increase of $p_{\text{deg,mRNA}i}$ by factor 5\\
  siRNA $i$ 100down & 100-fold down regulation of gene $i$ & increase of $p_{\text{deg,mRNA}i}$ by factor 100\\
  \end{tabular}
\end{ruledtabular}
 \caption{\label{tab:sres_poss_exps}Possible perturbations for the sensitivity design approach.} 
\end{table*}
\endgroup

The set of possible experiments for the benchmark model is summarized in Tab.\ \ref{tab:sres_poss_exps}. The total number of experiments is extended by allowing the application of two simultaneous perturbations on the system. Without double counting of needless perturbations on the same target, 442 possible combinations remain for perturbational experiments, including single perturbations. It is assumed that all species of the model can be measured individually and the measurement time points are chosen from the set $t \in \{ 1, 2, 3, 4,  6, 8, 10, 12 ,15, 20\}$. This yields a total number $\# {total} = \# {obs} \cdot \# {time} \cdot \#{perturb} = 12 \cdot 10 \cdot 442  = 53\,040$ of possible measurement points. Every of these 53\,040 data points corresponds to one row of a potential sensitivity matrix which contains the sensitivities of this measurement to all model parameters. All these rows of sensitivities for potential data points are merged together in the \textit{sensitivity matrix of all potential experiments}  $\mathcal{S}^{}_{pot}$.

For the illustration purpose, the measurement error $\sigma$ of the simulated data can be tuned arbitrarily, since repetition and averaging of the measurements of an experimental condition reduces the measurement error. Because $\sigma_{i}$ can be adapted individually for every data point, the used approximation for the Hessian
\begin{equation}
H \approx  \sum_{i,j} \frac{1}{\sigma_{i}} s_{i j} \cdot  \frac{1}{\sigma_{i}} s_{i j} = S_{\sigma}^\top S_{\sigma}\,
\end{equation}
can be adjusted for sensitivity matrices $S_{\sigma}$ whose $i$-th row is multiplied by one over the corresponding measurement error $\sigma_i$. To standardize the structure in the rows of the matrix with the sensitivities for all possible measurement points $\mathcal{S}^{}_{pot}$, all rows of $\mathcal{S}^{}_{pot}$ have been normalized by their Euclidean norm. This yields the matrix with \textit{normalized} sensitivities for all possible data points $\mathcal{S}^{norm}_{pot}$ with entries $s_{ij} \in [ -1, 1 ]$. The sensitivity matrix $S_{\mathcal{D}}$ for an arbitrary experimental design $\mathcal{D}$ in this setting is a subset of rows from the matrix $\mathcal{S}^{norm}_{pot}$. Thus, the experimental design process can be formulated as the selection of rows from $\mathcal{S}^{norm}_{pot}$.

A fundamental property of nonlinear ODE models is that the sensitivities depend on the estimated parameter values. This additionally hampers the construction of a desired structure in the sensitivity matrix, depending on the available data. However, in the asymptotic case of sufficient data the best fit on simulated data with an additive Gaussian noise will be close to the true parameters for the case of a sufficiently low noise level. Rather than performing the analysis on several simulated data sets from different realizations and averaging the results afterwards, the true parameter values are used throughout in order to facilitate the analysis and obtain general results. 

\subsubsection{Random designs}

As a reference, designs $\mathcal{D}$ are chosen randomly from $\mathcal{S}^{norm}_{pot}$. Fig.\ \ref{fig:random_designs} shows the resulting widths of the eigenvalue spectrum of 50 samples of designs $\mathcal{D}_{M}$ for different numbers of data points $M$. The random design approach yields a suboptimal design for a low number of data points $M$, whereas the width of the eigenvalue spectrum $w^\lambda$ decreases significantly for larger numbers of measurement points and yields matrices with an almost non-sloppy behavior for $M=1000$, i.e.\ $w^\lambda \approx 3$. The application of a huge number of perturbations on the system may thus diminish the effect of sloppiness. But this in turn implies that up to $1000$ independent data points from roughly the same number of different perturbative experiments have to be performed in order to describe a model with 29 dynamic parameters, which may exceed suitable limits in practice. 

\begin{figure}[htbp]
	\includegraphics[scale=.45]{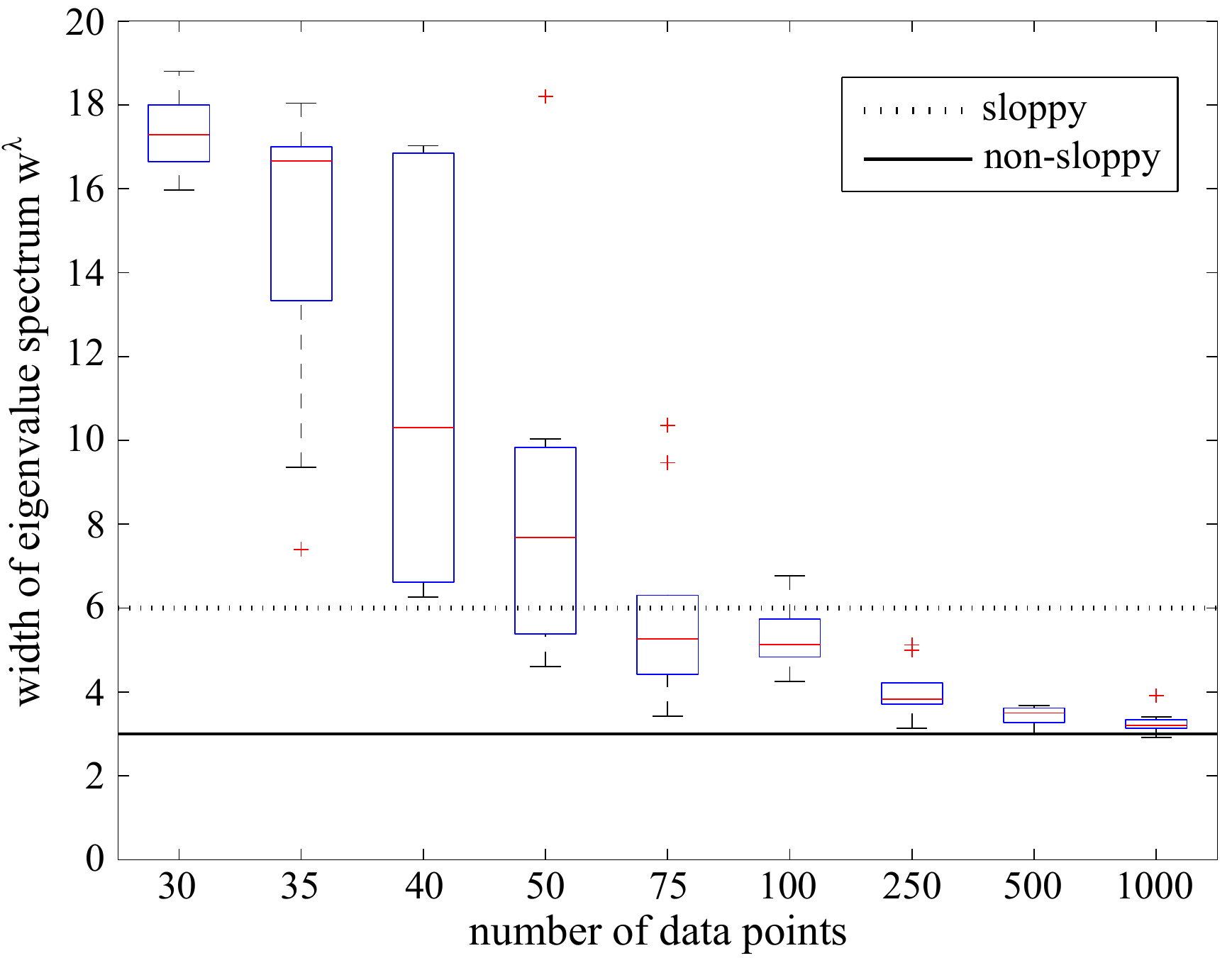}
	\caption{\label{fig:random_designs}(Color online) Box plot of spectral widths $w^\lambda$ of random designs for 50 samples of $S_\mathcal{D}$ for a various number of design points $M$. Samples with spectral width of the Hessian larger than $w^\lambda = 10^6$ are sloppy (dashed line), samples with width $w^\lambda \leq 10^3$ are non-sloppy (solid line).}  	
\end{figure}

Although this approach is able to remove the vertical correlation of densely sampled time course experiments by choosing only single measurement points from the time courses, the rows in the sensitivity matrices still exhibit recurrent structures in horizontal direction. For example, many measurements are barely sensitive for the parameters belonging to regulatory reactions in the benchmark model which were hard to constrain by data from appropriate experiments already in the setting of the original challenge. These restrictions from the model structure are the reason why the eigenvalue spectrum of the randomly chosen samples of $S_{\mathcal{D}}$ do not show the same behavior in terms $w^\lambda$ as the for the case of i.i.d.\ random entries described by the Mar\v{c}enkov-Pastur distribution.

\subsubsection{Exclusively sensitive experiments \label{sec:ex_sens}}

Apart from the random selection of rows from $\mathcal{S}^{norm}_{pot}$ specific rows can be chosen, in order to better assemble a structure of the sensitivity matrix which yields a small width of the eigenvalue spectrum. Due to normalization, the maximal absolute value of a component in the $i$-th row is $| s_{i\,m} |= 1$. Then in turn, all other components of this row are zero. If such an \textit{exclusively sensitive measurement} with $s_{ij}=1$ for $j=m$ and $s_{ij}=0$ for $j \neq m$ can be found for every estimated parameter $p_m$, a diagonal sensitivity matrix
\begin{equation}
S= \left( \begin{array}{cccc}
s_{11} & 0 & ... & 0 \\
0 & s_{22} & ... & 0 \\
\vdots & \vdots & \ddots & \vdots \\
0 & 0 & ... & s_{nn} \\
\end{array} \right) 
\label{eq_sens_100}
\end{equation}
could be constructed, with $s_{ii} = \pm 1$. Then, also the Hessian would have a diagonal form
\begin{equation}
H = S^{\top} S = \text{diag}(s_{11}^2, s_{22}^2, ...\, , s_{nn}^2)
\label{eq_100_hess}
\end{equation}
and the eigenvalues $\lambda_i$ of $H$ would be proportional to the squared nonzero sensitivities: 
\begin{equation}
\lambda_i(H) \sim \{ s_{ii}^2 \} \,\,\,,\, \text{where} \,\,\,\,\,\, s_{ii}^2 = 1\,\,\,\,\,\,\, i = 1, ... , n\,.
\end{equation}
Thus, all eigenvalues of the Hessian $\lambda_i = 1$ and the eigenspace of $H$ would be maximally degenerated. Furthermore, each row of $S$ and consequently every nonzero element $s^2_{ii}$ could be scaled by the corresponding adjustable measurement error $\sigma_i$ so that the eigenvalues of the Hessian are directly controlled by the measurement error. 

For the used benchmark model and experimental setup, such rows of \textit{exclusively sensitive experiments} in $\mathcal{S}^{norm}_{pot}$ do not exist for all parameters. Increasing the number of feasible experiments, e.g.\ by increasing the maximal number of combinations of perturbation, may reveal more of these \textit{exclusively sensitive experiments}. But due to the nonlinear character of the examined ODE models, the lack of analytic solutions for most systems and the dependence of the $s_{ij}(\vec{p})$ on the parameter values, a prediction of the sensitivities and the prediction of measurements showing such a structure in the sensitivities is impossible. Furthermore, this approach tends towards the case where single parameters can be measured directly and independently, which is a challenging demand in practice.

However, since sensitivities $s_{ij} = 1$ exist for some parameters in $\mathcal{S}^{norm}_{pot}$, a similar structure of $S$ is achieved by selecting the rows of $\mathcal{S}^{norm}_{pot}$ with the maximal absolute sensitivity $|s_{ij}|$ for each estimated parameter $p_j$. Fig.\ \ref{fig:100_percent} shows the results for this approach. Although the requested diagonal structure of the sensitivity matrix cannot be established entirely, the width of the spectrum of eigenvalues of the Hessian $w^\lambda = 2.11$ is clearly underneath the characteristic value $w^\lambda = 3$ which defines a \textit{non-sloppy} spectrum and is roughly one unit lower than for the best performing sample of random designs with $M = 1000$ data points in the previous section.

\begin{figure}[htbp]
	\includegraphics[scale=.35]{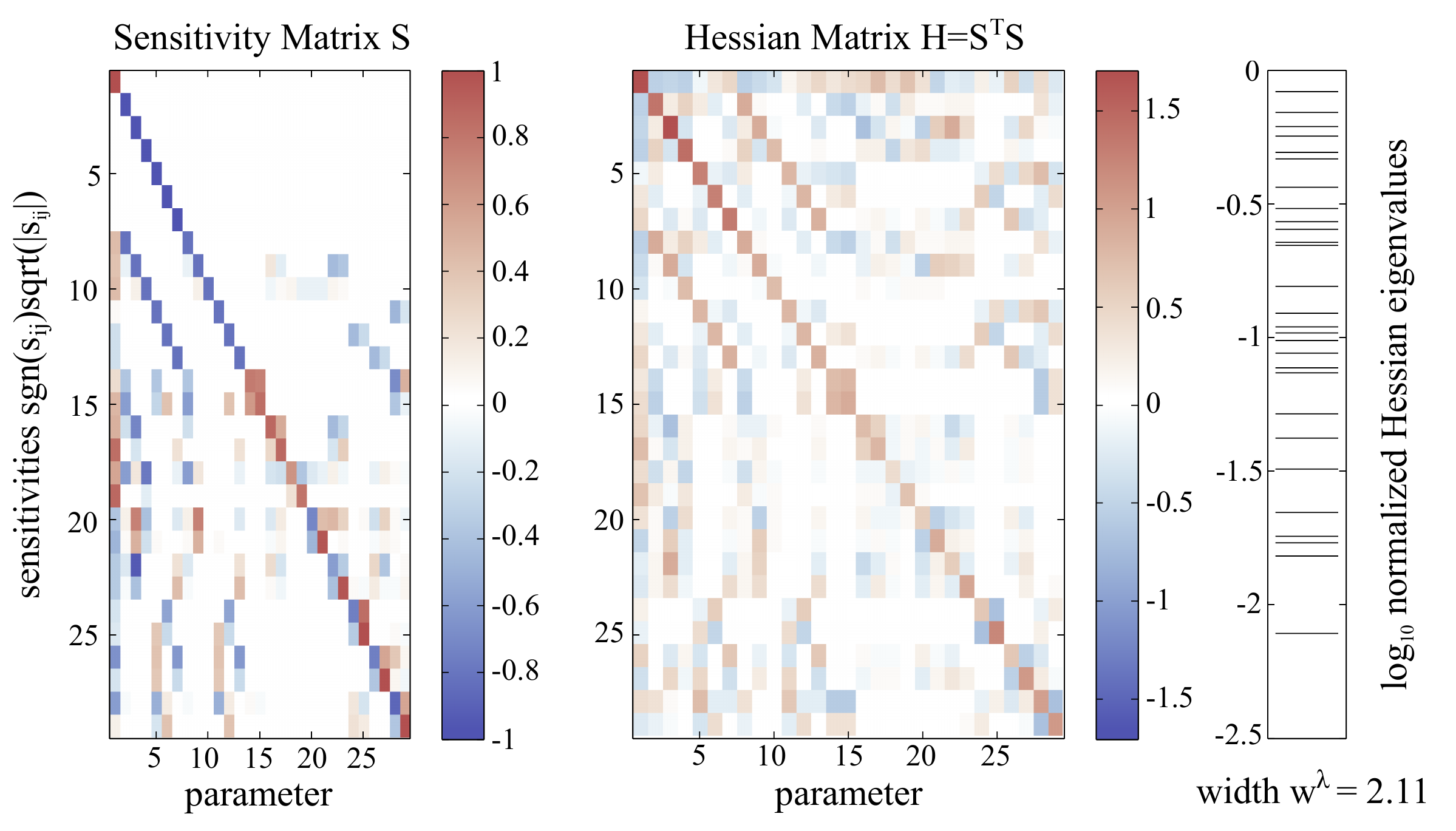}
	\caption{\label{fig:100_percent}(Color online) Sensitivity matrix $S$, Hessian matrix $H$ and logarithmic eigenvalue spectrum of $H$ for maximally sensitive experiments, i.e.\ with rows from $\mathcal{S}^{norm}_{pot}$ with maximal component $ s_{ij}$ for parameter $p_j$. For better illustration of small values, the square root with conserved sign, i.e.\ $s_{ij}^{\text{plot}} = \text{sgn}(s_{ij}) \cdot \sqrt{|s_{ij}|}$, is used for all shown matrix elements. Although exclusively sensitive measurements do not exist for all parameters, the width of the eigenvalue shrinks considerably to $w^\lambda = 2.11$ and uses only 29 data points.}  
			
\end{figure}

The width of an eigenvalue spectrum of a Hessian matrix with diagonal entries $s_{ii}^2 = 1$ as described by Eq.\ (\ref{eq_sens_100}) is $w^\lambda=0$. Consequently, the principal axes of the ellipsoid in the parameter space (cf.\ Eq.\ (\ref{eq_ellipsoid_symp})) would all have the same length and the principal axes would be aligned with the standard basis. If the eigenvectors of the sphere are rotated the Hessian matrix loses its diagonal form, whereas the the eigenvalue spectrum does not change. This in turn signifies, that also other structures in the sensitivity matrix yield a non-sloppy Hessian. In such a case, the rows in the sensitivity matrix $S$ differ from the special structure of the \textit{exclusively sensitive experiments}. Alhought such designs are more likely to be contained in $\mathcal{S}^{norm}_{pot}$, the selection of alternative designs remains a nontrivial task since it is not guaranteed that enough orthogonal rows can be found in $\mathcal{S}^{norm}_{pot}$.

\subsubsection{D-optimal search algorithm\label{sec_doptimal}}

The optimization criterion used to diminish the sloppiness effect, i.e.\ the minimization of $w^\lambda$, does not perfectly match with one of the classical optimization criteria derived from the Fisher Information Matrix, e.g.\ in \cite{atkinson_developments_1982}. The popular D-optimal criterion attempts to maximize the determinant of a matrix of the form $H =S^\top S$. Thus, a D-optimal design may also behave well in terms of the width of the normalized eigenvalue spectrum $w^\lambda$ although a it is not the optimal one in terms of non-sloppiness. 

\begin{figure}[!htbp]
	\includegraphics[scale=.47]{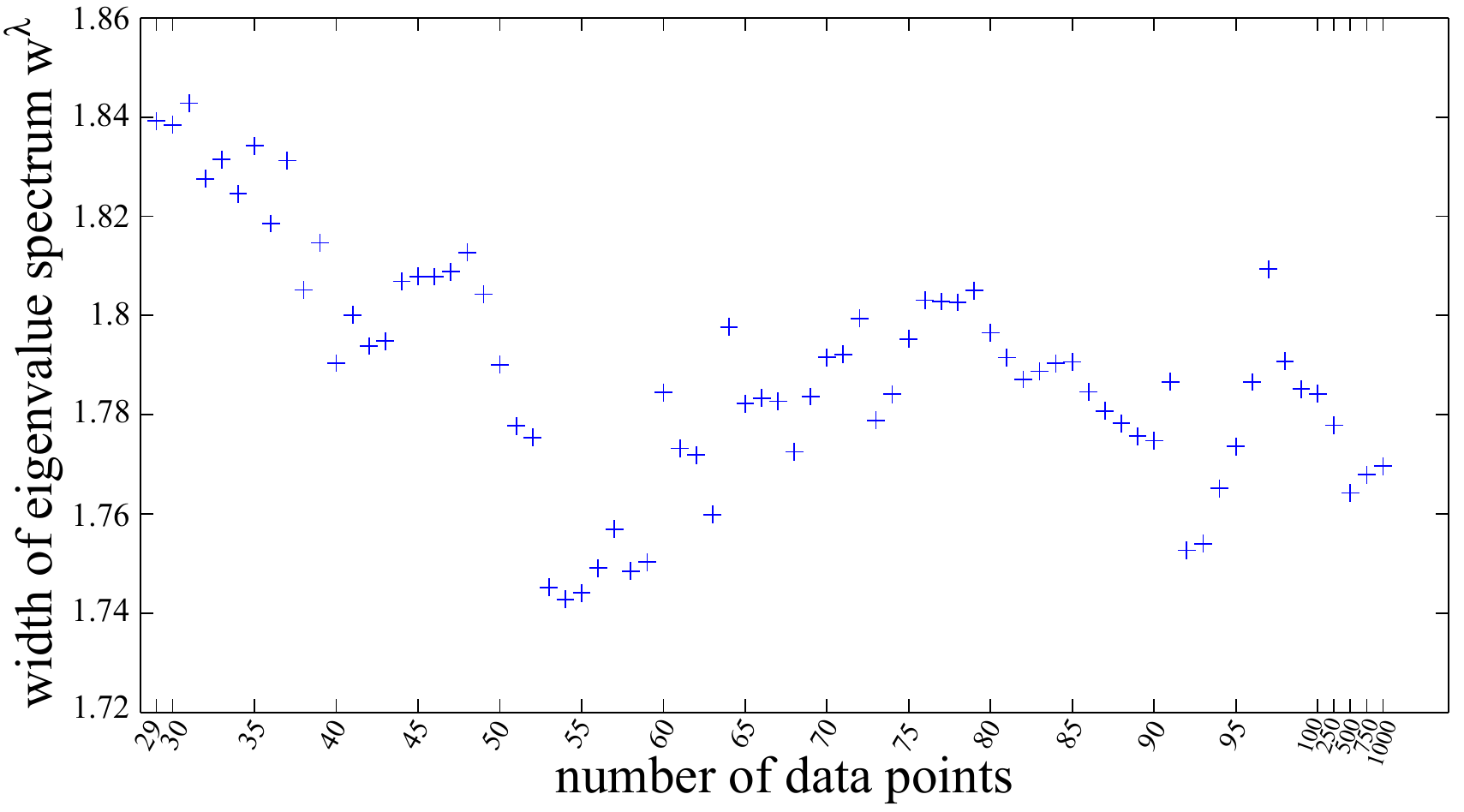}
	\caption{\label{fig:doptimal_plot}(Color online) Best performing designs, in terms of the spectral width $w^\lambda$, from 1000 suggested candidates of MATLAB's D-optimal search algorithm \texttt{chandexch.m} for different numbers of data points $M$. For every tested $M$, an undoubtedly non-sloppy design with $w^\lambda \leq 1.85$ is found.}  
\end{figure}

\begin{figure}[!htbp]
	\includegraphics[scale=.32]{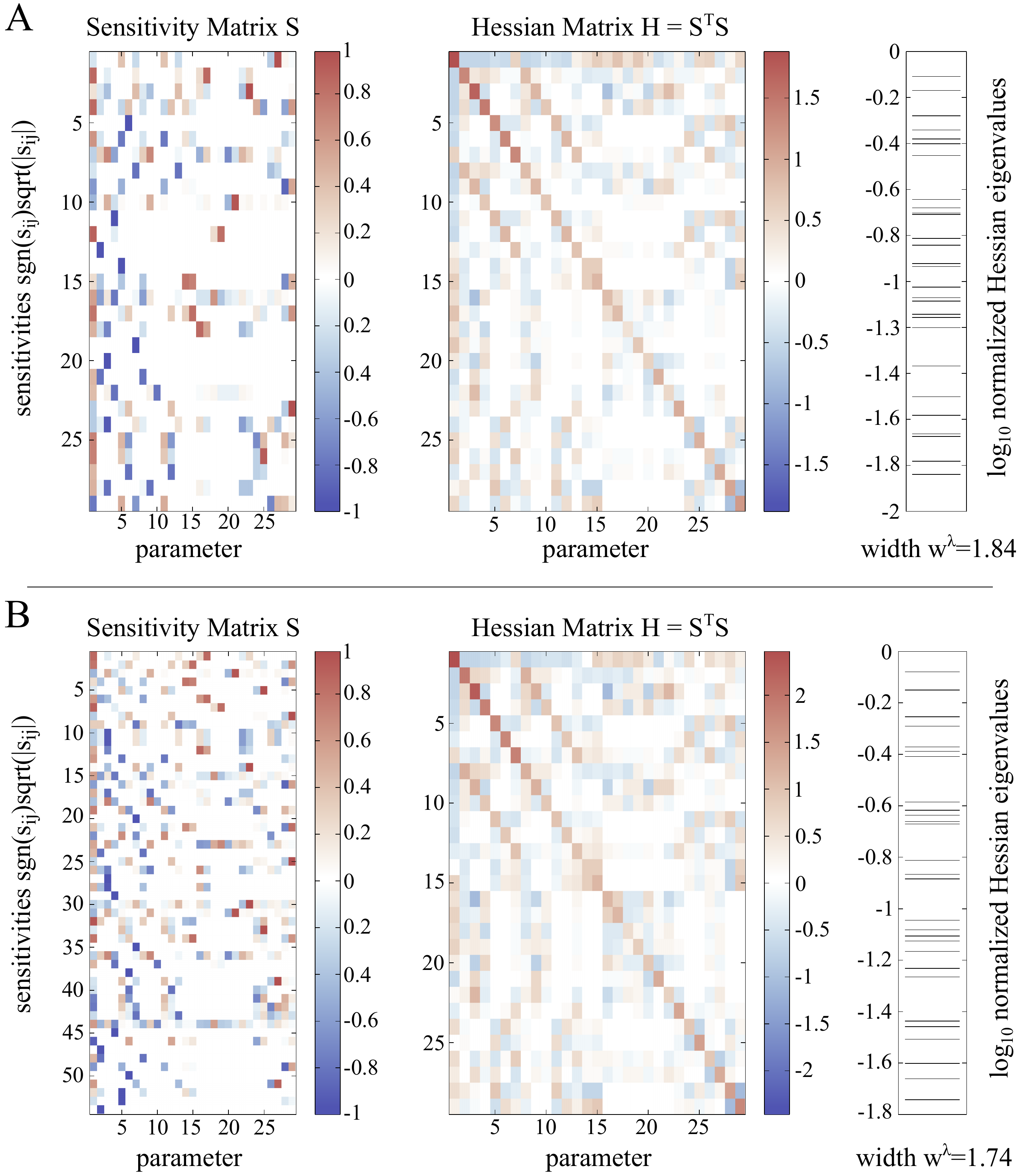} 
	\caption{\label{fig:doptimal_examples}(Color online) Best designs, in terms of the spectral width $w^\lambda$, from D-optimal candidates after 1000 runs of \texttt{chandexch.m}. For better illustration of small values, the square root with conserved sign is used for all shown matrix elements. Panel A shows the design for the minimal number of data points $M = 29$ with spectral width $w^\lambda = 1.84$, which is even less than for the \textit{exclusively sensitive approach}, cf.\ Fig.\ \ref{fig:100_percent}. Panel B presents the overall best design of all investigated $M$ values for $M^\ast = 54$ with clearly non-sloppy spectral width $w^\lambda = 1.74$. The sensitivity matrix exhibits only few regularities and is almost unstructured, but only sparsely filled.} 
\end{figure}

In the following, an algorithm for identifying D-optimal designs is utilized to suggest possible candidates for optimal designs which are then tested in terms of sloppiness. Here, we use the implementation of the function \texttt{chandexch.m} from the statistics toolbox of MATLAB as a D-optimal search algorithm. Fig.\ \ref{fig:doptimal_plot} shows the widths of the eigenvalue spectrum of the Hessian for the best performing design from 1000 runs of the D-optimal search algorithm for each number of data points $M$. For every tested number of data points, an optimal design  chosen from the suggested D-optimal candidates which yields a spectral width $w^\lambda \leq 1.85$. This signifies that the used D-optimal search algorithm is able to find a non-sloppy design of the sensitivity matrix $S$ for each analyzed number of data points $M$. It also selects the measurement points for the minimal number of data points $M=29$ superior compared to the design from the \textit{mostly exclusively sensitive experiments} structure approach. This best performing design for $M=29$ is presented in the upper panel of Fig.\ \ref{fig:doptimal_examples}. The lower panel of Fig.\ \ref{fig:doptimal_examples} shows the overall best performing design with $M = 54$ data points and spectral width $w^\lambda = 1.74$. This clearly demonstrates that the eigenvalue characteristics of the Hessian of ODE models can be controlled by the choice of an appropriate experimental design and therefore sloppiness is no general characteristics of ODE models.

\section{Conclusion \label{conclusion}}

In Chap.\ \ref{chap:cause}, it was shown that the stepwise introduction of correlations and structures in the sensitivity matrix, motivated by the properties of the investigated ODE models and the applied experimental methods, yields a gradually increasing width of the Hessian eigenvalue spectrum. By the adjustment of the coefficients of the processes generating the structure in the sensitivity matrix, the eigenvalue spectrum of the Hessian can be tuned to almost arbitrary widths. The inclusion of insensitive experiments for certain parameters, imitated by randomly chosen empty entries in the sensitivity matrix, leads to a spectrum with a likewise large eigenvalue spread but the eigenvalues are more equidistantly distributed which is a property of \textit{sloppy} eigenvalue spectra, like published in \cite{gutenkunst_universally_2007}. Especially, the block structure of highly correlated rows of the sensitivity matrix which are typically obtained in applications for time course measurements has a huge impact on the spectral width. 

Understanding the origins of a sloppy eigenvalue spectrum enables experimental design considerations, in order to reduce the sloppiness effect. One result from the analysis of random matrices is that vertical correlations in the sensitivity matrix rapidly increase the spectral width. Thus, dense time sampling of measurement points compared to the time scales of the underlying dynamics yields a suboptimal design in terms of sloppiness. Such a vertical structure in the sensitivity matrix can be avoided by selecting only characteristic data points which results in a reduction of the correlation of entries in the sensitivity matrix. 

Furthermore, also the structure in horizontal direction of a sensitivity matrix should be minimized, so that this structure converges to the case of completely independent entries. This would lead to a decrease of the width of the eigenvalue spectrum to a natural limit, as described by the Mar\v{c}enkov-Pastur distribution. However, the horizontal structure is predetermined completely by the model structure, since specific connections in the network yield dependencies of the internal states on the variation of certain parameters. 

The manipulation of these model structures, e.g.\ by short interfering RNA (siRNA) mediated knockdown experiments \cite{moazed_small_2009}, can be realized by an appropriate selection of measurements which induce the desired structure in the sensitivity matrix. However, this requires a preferably large amount of possible experiments in order to have access to complementary structures of the sensitivities. Nevertheless, the optimal selection of measurements remains a nontrivial task which demands for appropriate experimental design methods.

The analysis of randomly chosen designs yields acceptable results in Chap.\ \ref{chap:cure} for designs width large numbers of measurements. This result is related to the fact that the time points are selected in a way that almost every data point originates from another experimental perturbation. Although, these huge numbers of independent experiments would not be feasibly in practice, the block-wise structure of correlation from densely sampled data from time course experiments was avoided. The construction of a sensitivity matrix by experiments which are almost exclusively sensitive to single parameters utilizes the characteristic structure in the sensitivities for a specific experiment and yields good results for a small number of data points, i.e.\ the number is comparable with feasible limits in an application setting. However, due to the typical network structure of the investigated systems, the experimental realization of a specific structure in terms of required perturbations in the sensitivities is challenging. Especially for networks with a pathway structure, measurements of downstream located species will always depend on the upstream reactions and consequently on the parameters controlling these upstream reactions. To isolate the behavior of such a downstream compartment, e.g.\ complete knockouts of all connections to the upstream compartments have to be performed. Furthermore, these special perturbations have to be manageable experimentally.

It should be noted at this point, that this complexity of feasible perturbations is only required for minimization of the width of the eigenvalue spectrum. It is not necessary for identification of the parameter components in other concepts to assess the uncertainties of parameters or in terms of identifiability, as presented in e.g.\ in \cite{raue_structural_2009}. The width of the normalized eigenvalue spectrum only represents the ratio of the smallest and largest width of the ellipsoid from Eq.\ (\ref{eq_ellipsoid_symp}) along its principal axes. Its orientation in the parameter space or its projection on the bare parameter axes, which is of interest for confidence intervals of the point estimatesm, is not considered in the sloppiness discussion, but may be more relevant in applications. Moreover, the whole analysis is restricted to effects of maximal second order derivatives of $\chi^2$, which can be misleading due to the nonlinearity of the investigated models, especially in the case of insufficient amount of data. 

For the benchmark model and experimental setup, we could show that with the use of the D-optimal search algorithm it is possible to select measurement points yielding a Hessian with a minimal spectral width $w^\lambda = 1.74$, which is a value far away from a sloppy Hessian and even significantly lower than the non-sloppy threshold, as defined e.g.\ in \cite{brown_statistical_2003, waterfall_sloppy-model_2006}. Thus, it could be shown that it is possible by optimal experimental design methods to diminish the sloppiness effect to a minimum. Furthermore, the best design was neither a complete measurement of all accessible observables nor a direct measurement of single parameters as often surmised, but consists of single measurements with nonzero sensitivities for multiple parameters.

Our results suggest that sloppiness in dynamical models and fundamental physics of diffusion and Ising models as reported in \cite{machta_parameter_2013} is merely a phenomena of coincidence than a common cause. We discussed that the model structure influences the structure of the sensitivity matrix promoting the appearance of sloppiness, but the intensity of this effect is controlled by the design of experiments, i.e.\ by the data. Thus, assigning sloppiness to a model as a general characteristic is incomplete without discussing experimental design aspects. The sloppiness issue, as typically discussed in the literature, e.g.\ in \cite{gutenkunst_universally_2007}, only refers to a standard design where all dynamic variables are observable as a continuous time course. Although the discovery of a non-sloppy design may be still challenging since the sensitivities depend on the unknown parameters, the strong dependency of the eigenvalue spectrum of the Hessian on the experimental design weakens the declaration of sloppiness as a universal property of ODE \textit{models}.

\begin{acknowledgments}

The authors would like to thank D.A.\ Rand for helpful discussions. This work was supported by the grants 0315766-VirtualLiver and 0316182A-SBEpo of the German Federal Ministry of Education and Research (BMBF).

\end{acknowledgments}

\bibliography{CauseAndCureOfSloppinessInODEModels}

\end{document}